\documentclass[11pt,preprint]{aastex}

\usepackage{amsmath}
\usepackage{amssymb}
\usepackage{mathbbol}
\usepackage{dsfont}

\shorttitle{ELVIS' discs aren't turning}
\shortauthors{Pawlowski et al.}

\begin{document}

\title{Co-orbiting planes of sub-halos are similarly unlikely\\ around paired and isolated hosts}

\author{Marcel S. Pawlowski and Stacy S. McGaugh}
\affil{Department of Astronomy, Case Western Reserve University,\\
              10900 Euclid Avenue, Cleveland, OH, 44106, USA}
\email{marcel.pawlowski@case.edu}

\begin{abstract}
Sub-halos in dark-matter-based cosmological simulations tend to be distributed approximately isotropically around their host. The existence of highly flattened, co-orbiting planes of satellite galaxies has therefore been identified as a possible problem for these cosmological models, but so far studies have not considered the hosts' environments. That satellite planes are now known around both major galaxies in the Local Group raises the question whether they are more likely around paired hosts. In a first attempt to investigate this possibility we focus on the flattening and orbital coherence of the 11 brightest satellite galaxies of the vast polar structure (VPOS) around the Milky Way (MW). We search for VPOS-analogues in the ELVIS suite of cosmological simulations, which consist of 24 paired and 24 isolated host halos. 
We do not find significant differences between the properties of sub-halo distributions around paired and isolated hosts.
The observed flattening and the observed orbital alignment are each reproduced by only 0.2 to 2 per cent of paired and isolated systems incorporating the obscuration of satellites by randomly oriented galactic discs. 
Only one of all 4800 analysed realisations (0.02 per cent) reproduces both parameters simultaneously, but the average orbital pole of this sub-halo system does not align as well with the normal to the plane fit as observed.
That the MW is part of a galaxy pair thus does not help in explaining the existence of the VPOS if the satellite galaxies are identified with sub-halos found in dissipationless simulations.
\end{abstract}


\section{Introduction}

The distribution of satellite galaxies around the Milky Way (MW) is highly anisotropic: they align in a narrow plane perpendicular to the Galactic disc \citep{LyndenBell1976,Kroupa2005}.
Many globular clusters and streams in the MW halo are part of the same structure, which has been termed the vast polar structure  \citep[VPOS, ][]{Pawlowski2012a}.
While our knowledge of fainter objects is affected by the uneven sky coverage of surveys in which they are detected, the 11 brightest ('classical') satellites are generally believed to be a less biased distribution. Proper motion measurements reveal that eight of them are consistent with co-orbiting in the VPOS \citep{Pawlowski2013b}.

This phase-space correlation of the MW satellites is difficult to reconcile with expectations based on the current standard model of cosmology. Dark matter sub-halos in cosmological simulations do not show the observed degree of coherence. Claims of consistency with simulations \citep[e.g.][]{DOnghia2008,Li2008,Libeskind2009,Deason2011,Lovell2011,Wang2013,Bahl2014} have been found not to hold in view of additional observational data \citep{Metz2009,Pawlowski2012b,Pawlowski2013b} or to be based on flawed analyses \citep{Ibata2014,Pawlowski2012b,Pawlowski2014b}.

The possibility that the environment of a host affects the chance to find correlated sub-halo planes has not yet been investigated. The MW is part of the Local Group (LG), together with M31. The discovery of an apparently rotating satellite plane around M31 \citep{Ibata2013} and of two planes containing almost all isolated dwarf galaxies in the LG, the dominant one of which aligns with Magellanic Stream which is part of the VPOS \citep{Pawlowski2012a,Pawlowski2013a}, indicate that the nearby environment is possibly related to the satellite structures \citep[see also][]{Pawlowski2014}.

The high-resolution cold dark matter simulations of the 'Exploring the Local Volume in Simulations' (ELVIS) project \citep{GarrisonKimmel2014} offer an opportunity to test whether satellite planes are more likely to be present around paired hosts. Half of its 48 host halos are in a paired configuration, while the other half are isolated but matched in mass.
We use this dataset to test whether the probability to find VPOS-like satellite planes among the 11 most-massive satellites is different for paired and isolated hosts and whether being part of a paired group affects the distribution of the 11 to 99 most-massive satellites. 
Sect. \ref{sect:method} summarized the simulations, sample selection and analysis, Sect. \ref{sect:results} presents our results and conclusions are drawn in Sect. \ref{sect:conclusion}.


\section{Method}
\label{sect:method}

The ELVIS suite \citep{GarrisonKimmel2014} is a set of cosmological zoom simulations focussing on 12 pairs of main halos with masses, separations and relative velocities similar to those of the MW and M31. A control sample of 24 isolated halos matching the paired ones in mass has also been simulated. The simulations are dissipationless ('dark-matter-only'), based on WMAP-7 cosmological parameters \citep{Larson2011} and complete for sub-halo masses down to $\approx 10^7 M_{\sun}$, thus resolving objects comparable to the classical MW satellites.

We use the publicly available present day ($z = 0$) halo catalogues. For the case of paired hosts, both sub-halo systems are analysed in the same way. To be comparable to the MW satellite system, only sub-halos between 15 and 260\,kpc from the center of their host are considered. 
They are ranked by stellar mass, as determined by abundance matching (AM) applied to the maximum mass $M_{\mathrm{peak}}$\ they had over their history. While different AM prescriptions, such as the \citet{Behroozi2013} model or the preferred model by \citet{GarrisonKimmel2014}, differ in the stellar mass assigned to a given $M_{\mathrm{peak}}$, they preserve the $M_{\mathrm{peak}}$-ranking of satellites. As we select the highest-ranked satellites (the absolute mass is not part of the selection) the different prescriptions result in the same selection.

For the comparison to the observed VPOS the analysis accounts for the obscuration of satellites by the MW. All sub-halos within $\pm 11.5^{\circ}$\ from an obscuring disc (corresponding to 20\% of the sky) are ignored. For each of the 48 sub-halo systems 100 realisations are generated by drawing the top 11 sub-halos from outside the obscured area of different randomly oriented obscuring discs.

For each realisation a plane is fitted to the 11 sub-halos. The method is identical to the one applied to the observed satellite positions \citep{Pawlowski2013a} and the Millennium-II simulation \citep{Pawlowski2014b}. It finds the principal axes of the satellite distribution by determining the eigenvectors of the moments of inertia tensor constructed using non-mass-weighted positions. The orientation of the best-fit plane is described by its normal vector and the following parameters describing the shape of the distribution are determined (parameters measured for the MW satellites using positions compiled by \citet{McConnachie2012} are given in brackets):
\begin{itemize}
 \item $r_{\mathrm{per}}$, the root-mean-square (RMS) height of the sub-halos perpendicular to the plane ($r_{\mathrm{per}}^{\mathrm{obs}} = 19.6$\,kpc).
 \item $r_{\mathrm{par}}$, the RMS radius of the sub-halos projected into (parallel to) the best-fit plane measured from the center of their host ($r_{\mathrm{par}}^{\mathrm{obs}} = 129.5$\, kpc).
 \item $c/a$\ and $b/a$, the short- and intermediate-to-long RMS axis ratios, respectively ($(c/a)^{\mathrm{obs}} = 0.182$\ and $(b/a)^{\mathrm{obs}} = 0.508$).
\end{itemize}

Eight of the 11 MW satellites have orbital poles which align with the normal of the VPOS, indicating that these satellites co-orbit within the structure. We consider this essential property of the VPOS by measuring the following two parameters and comparing them to those determined for the observed MW satellites using the same method as \citet{Pawlowski2013b}:
\begin{itemize}
 \item $\Delta_{\mathrm{std}}$, the spherical standard deviation of the eight most-concentrated orbital poles which measures their concentration ($\Delta_{\mathrm{std}}^{\mathrm{obs}} = 29.3^{\circ}$).
 \item $\theta_{\mathrm{VPOS}}$, the angle between the average direction of the eight most-concentrated orbital poles and the normal defining the best-fit plane, which measures the alignment with the plane ($\theta_{\mathrm{VPOS}}^{\mathrm{obs}} = 18.9^{\circ}$).
\end{itemize}

The 48 hosts are split up into two classes: 20 paired and 24 isolated halos. Like \citet{GarrisonKimmel2014} we exclude the halos Serena \& Venus and Siegfried \& Roy from the paired sample because unlike the LG these contain a third massive halo at a distance of about 1\,Mpc. In addition, for each realisation a randomized system having the same radial distribution as the simulated systems is generated by rotating each sub-halo into a random direction before applying the obscuration cut. These show which properties are to be expected for isotropic systems.


\section{Results}
\label{sect:results}

\subsection{Searching VPOS-analogues}
\label{sect:VPOS}

\begin{figure*}
   \centering
   \includegraphics[width=80mm]{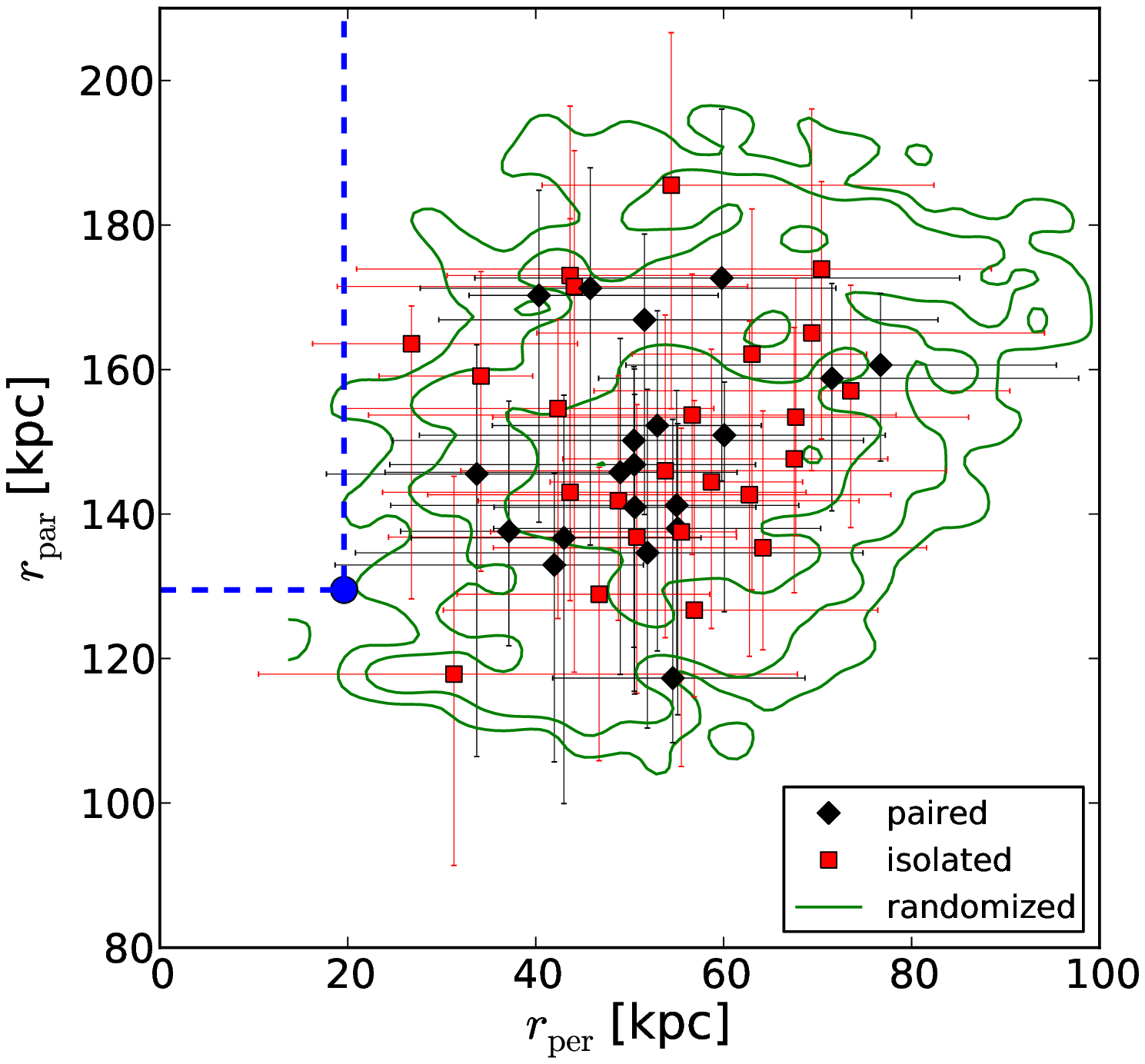}
   \includegraphics[width=80mm]{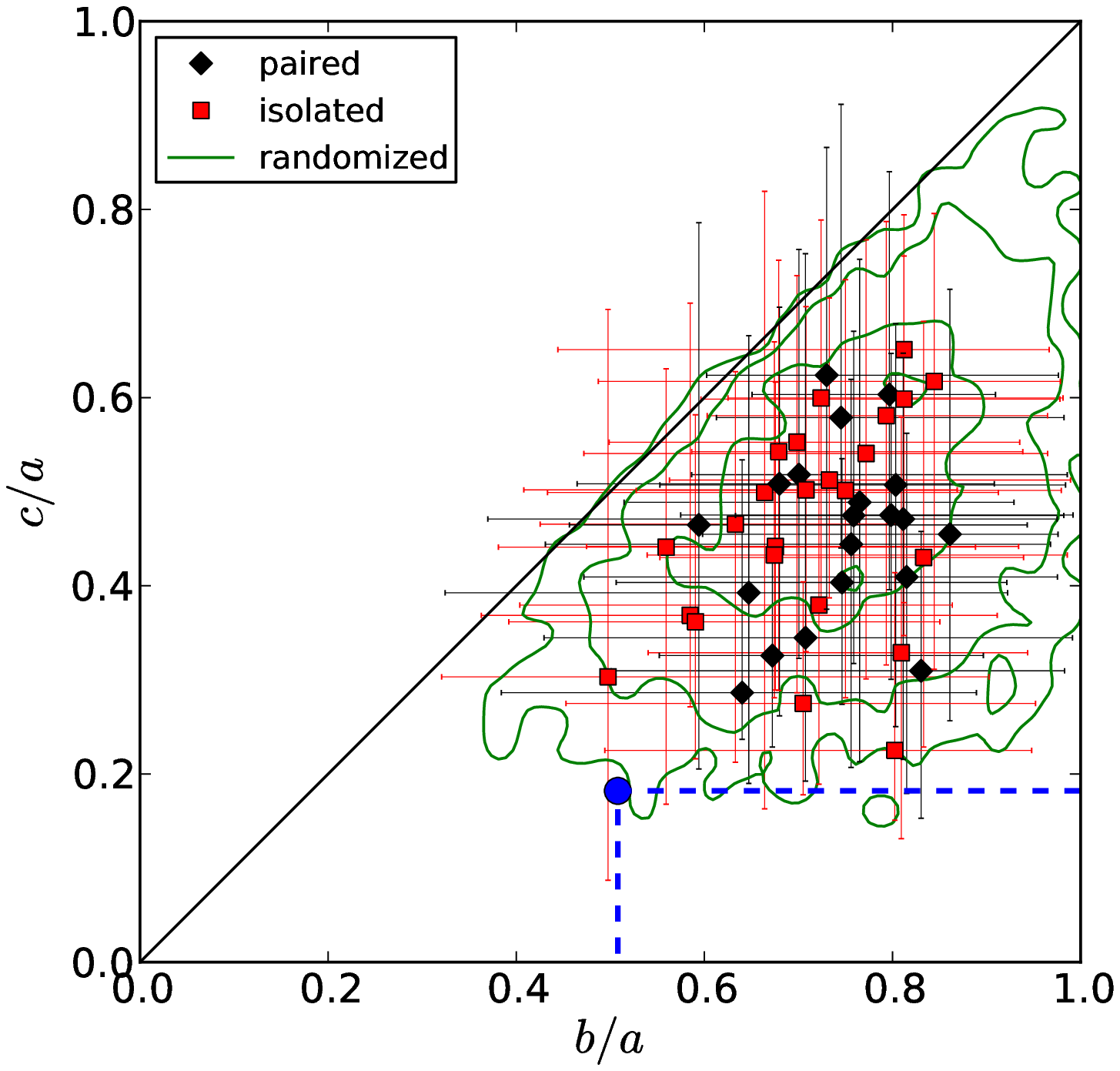}
   \includegraphics[width=80mm]{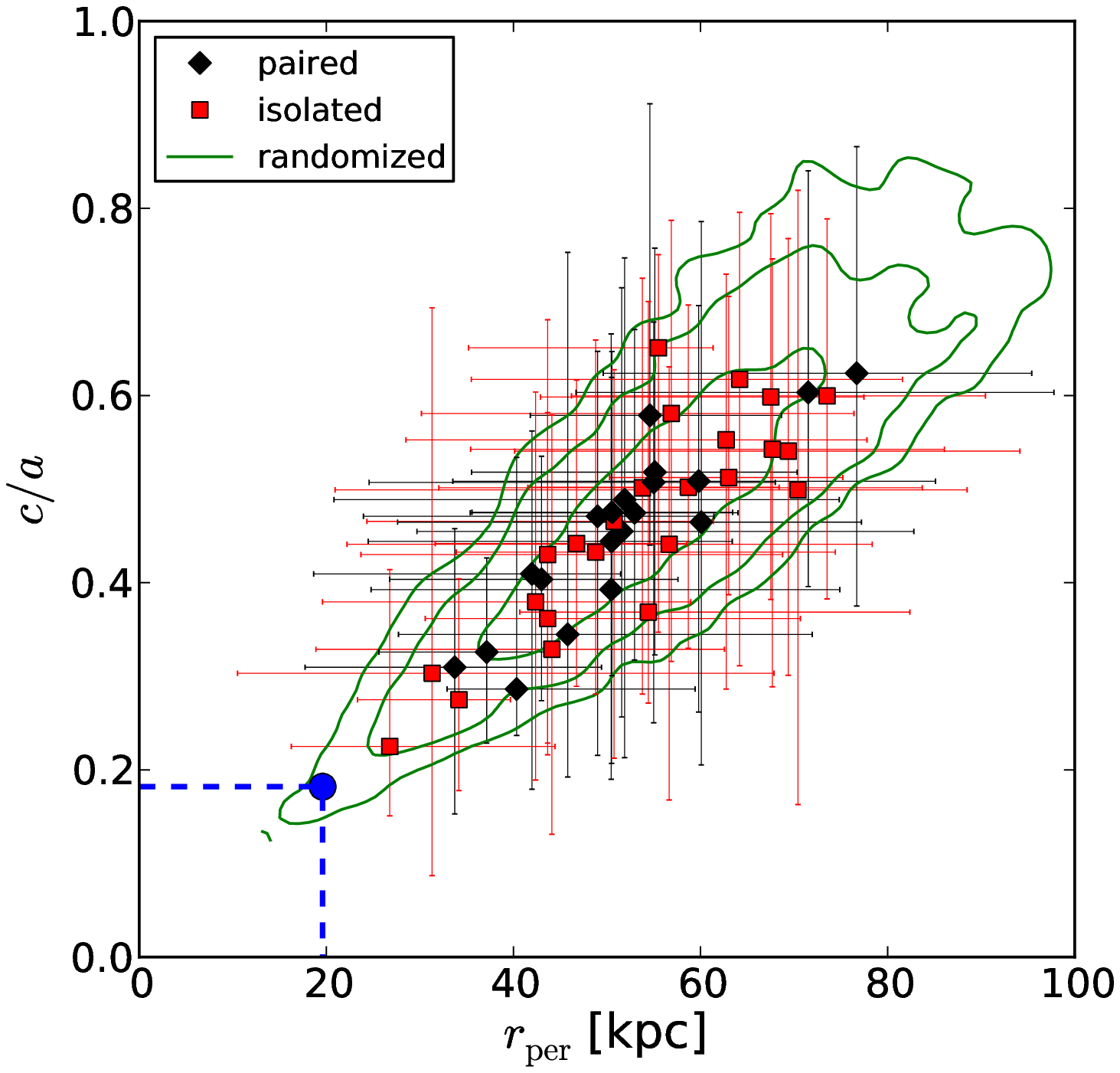}
   \includegraphics[width=80mm]{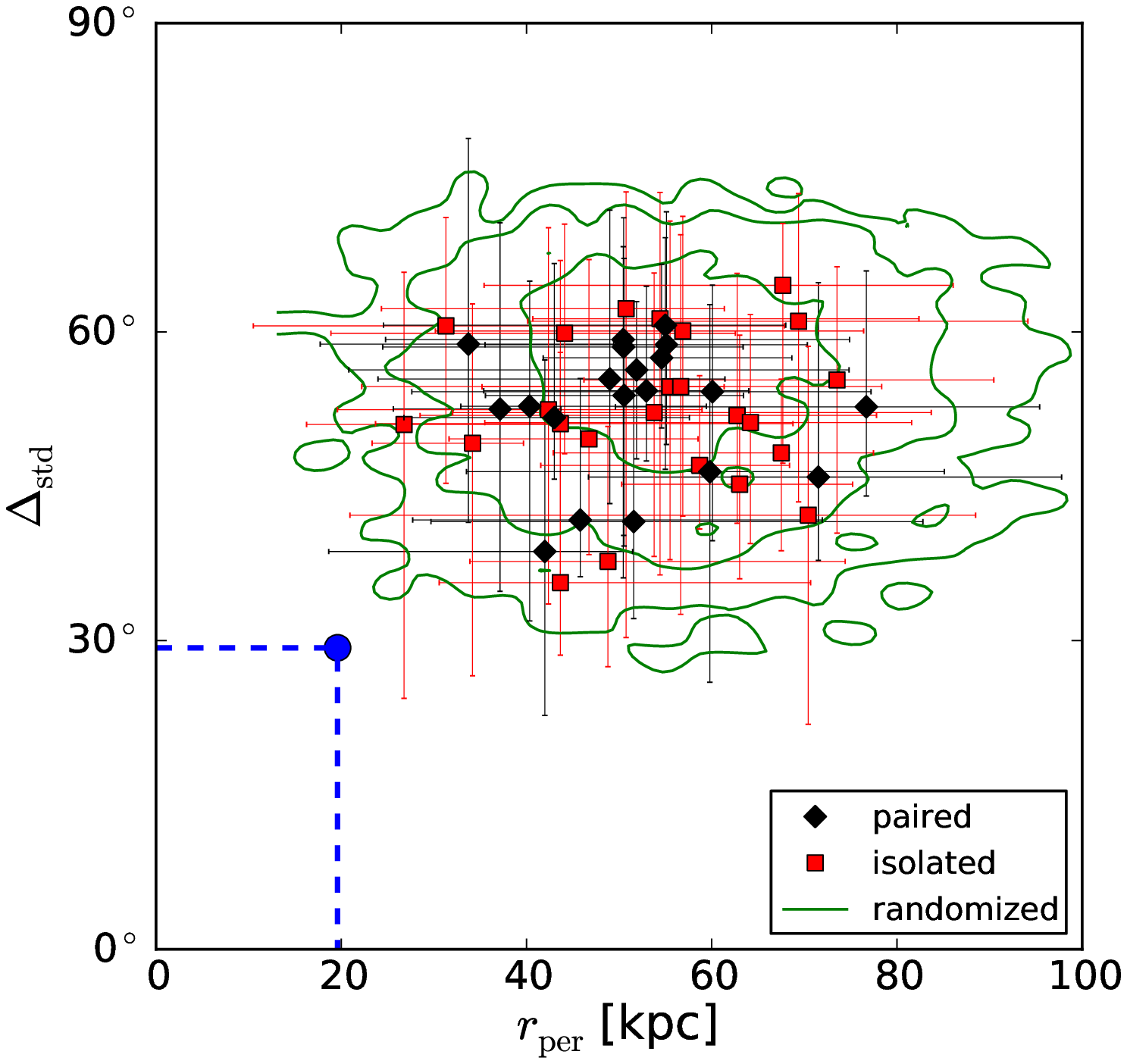}
   \caption{
   Median (symbols) and maximum and minimum values (error bars) for the disc-fit parameters determined from 100 random obscuring disc realisations for each of the paired and isolated hosts. The blue dot gives the parameters determined from the 11 most-luminous MW satellites. Models within the areas marked by the dashed lines reproduce these VPOS properties. The green contours contain 50, 90 and 95\% of all randomized realizations. 
   }
              \label{fig:scatter}
\end{figure*}

\begin{figure*}
   \centering
   \includegraphics[width=80mm]{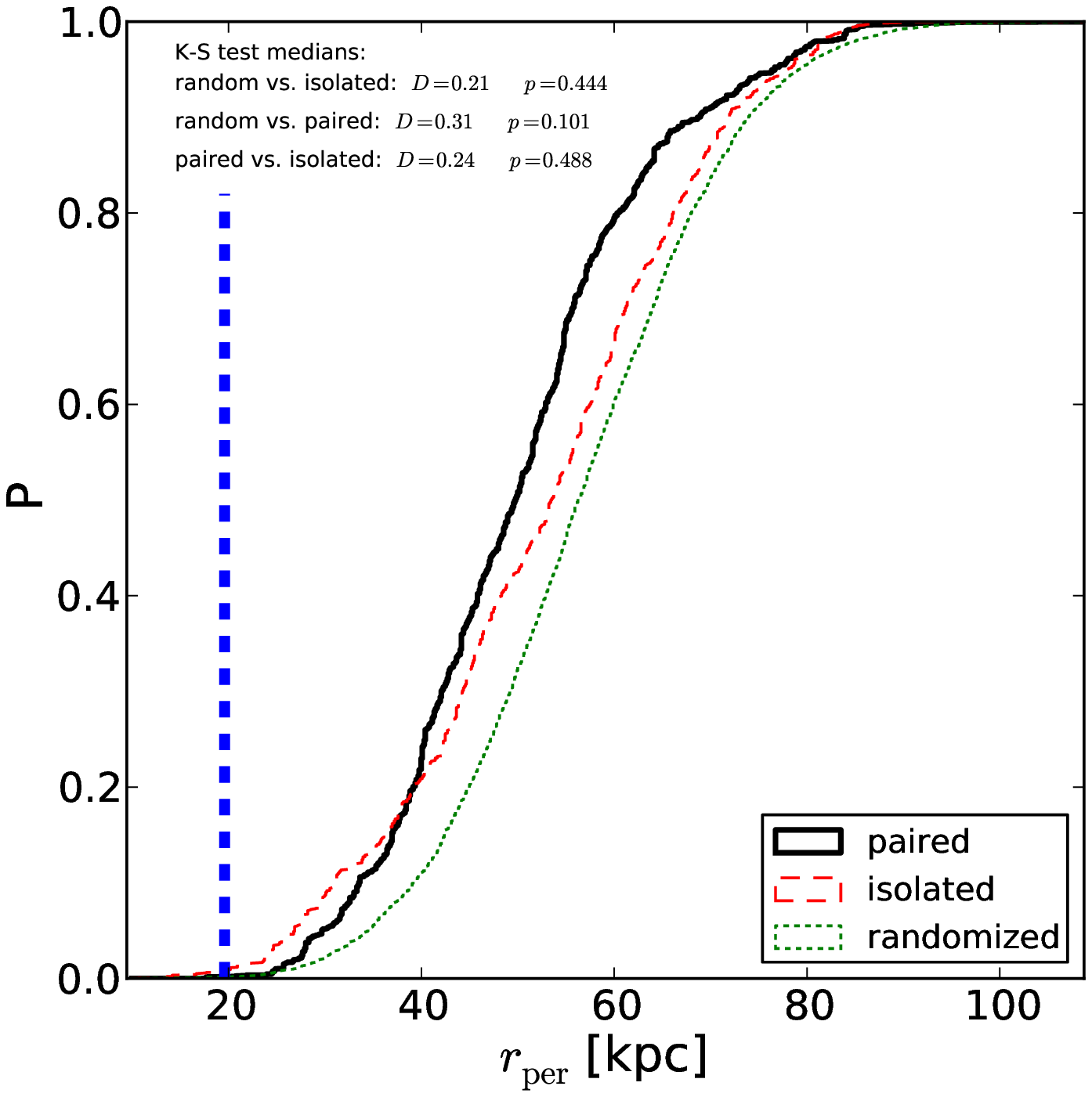}
   \includegraphics[width=80mm]{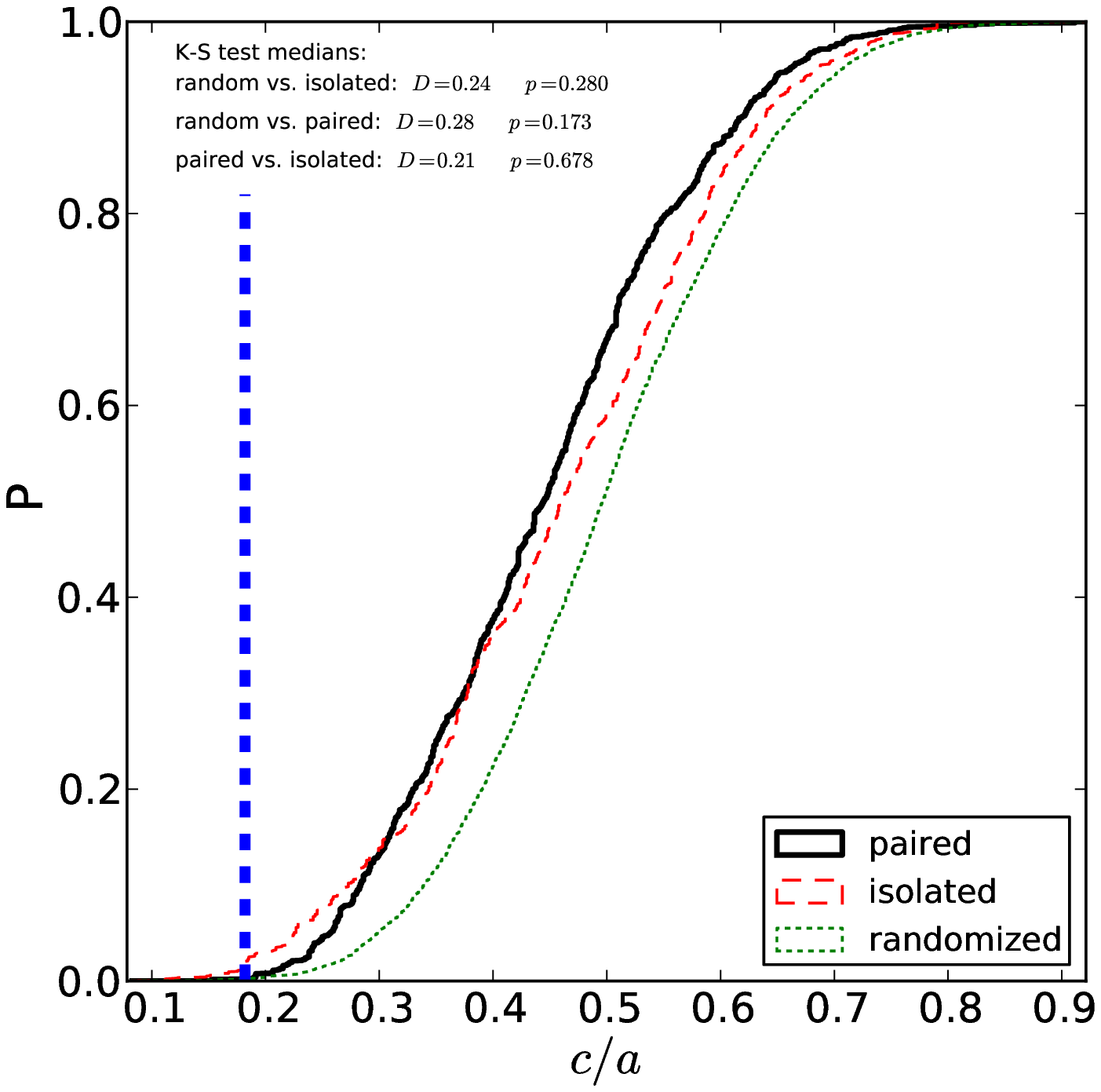}
   \includegraphics[width=80mm]{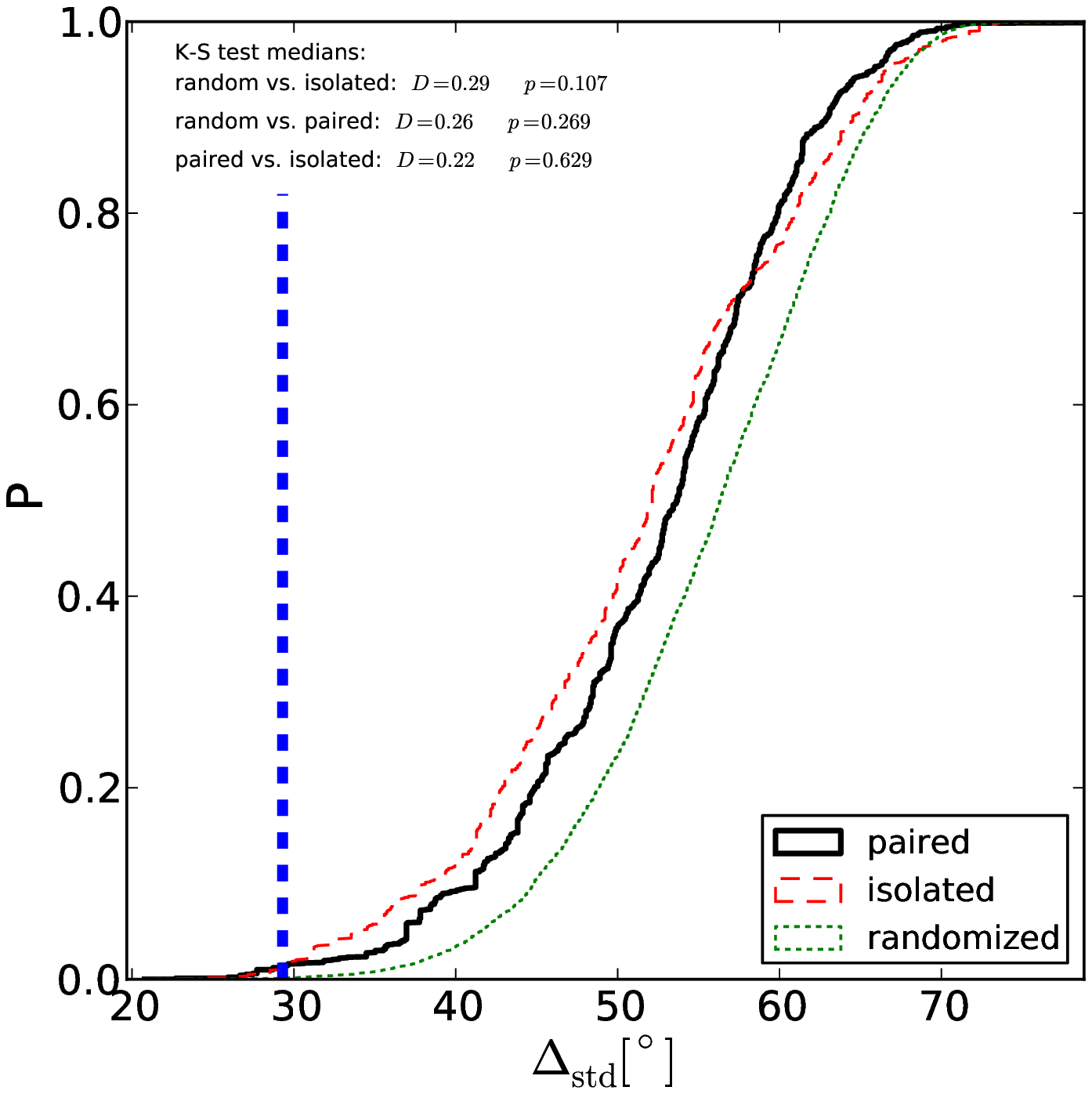}
   \includegraphics[width=80mm]{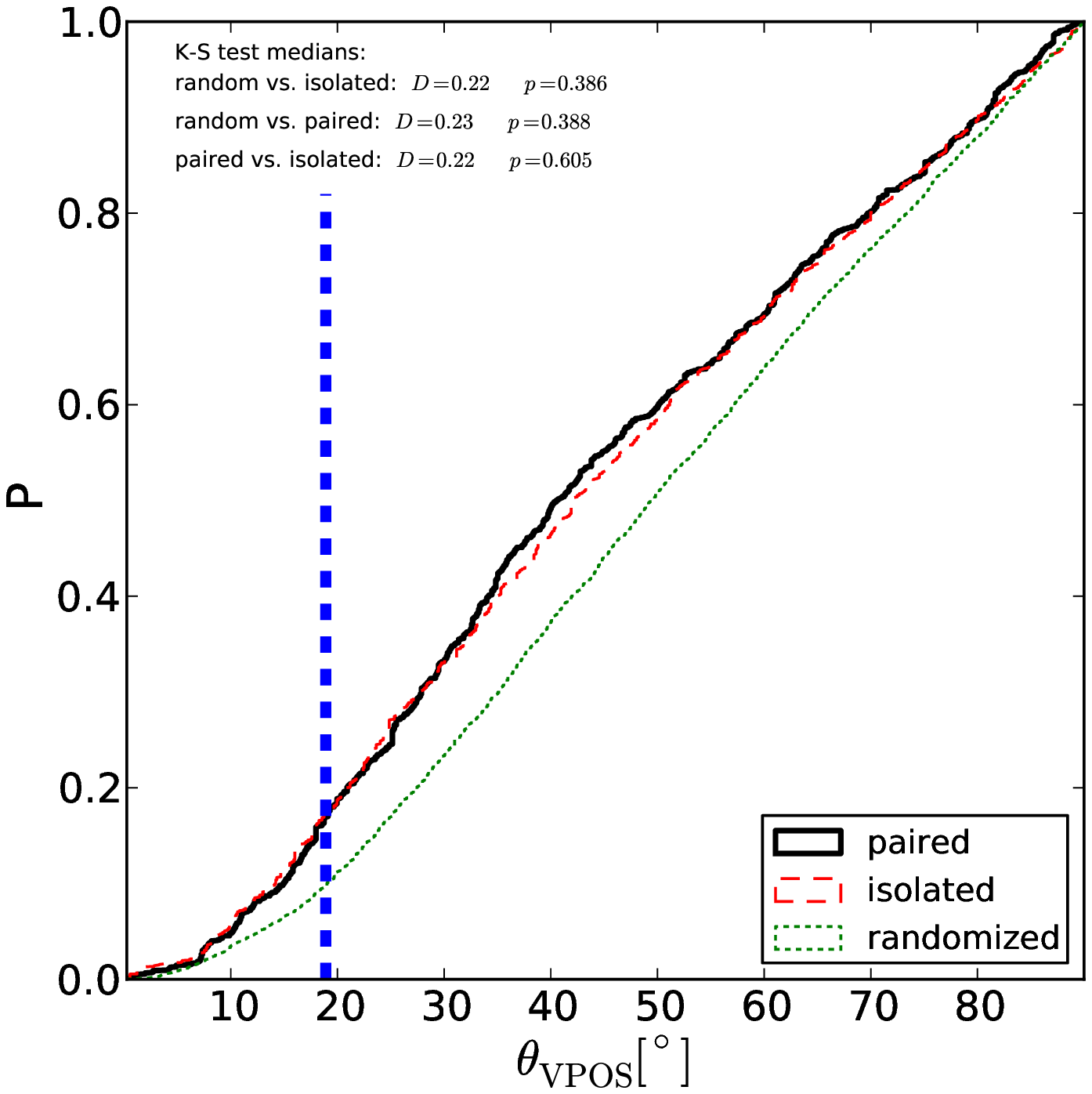}
   \caption{
   Cumulative distributions of $r_{\mathrm{per}}$, $c/a$, $\Delta_{\mathrm{std}}$\ and $\theta_{\mathrm{VPOS}}$. The blue dashed line indicates the respective value measured for the VPOS.
   Each halo system contributes 100 realisations with randomly oriented obscuring discs to the cumulative curves, which smooths the distributions, but the different realisations for one host are not independent. Therefore 100 Kolmogorov-Smirnov-tests have been performed for combinations containing only one realisation per host and the median parameters of these tests have been determined. They show that with the current number of hosts is is not possible to rule out the null-hypotheses that the paired, isolated and randomized distributions have been drawn from the same parent distribution.
   }
              \label{fig:cumulative}
\end{figure*}

To be similarly correlated in phase-space as the observed MW satellites, a simulated system has to have at least as extreme plane-fit parameters, which results in the criteria compiled in Table \ref{tab:VPOSresults}. It also lists which fractions of realisations fulfil the criteria. The distribution of the plane-fit parameters are shown in Figures \ref{fig:scatter} and \ref{fig:cumulative}.

As apparent from Fig. \ref{fig:scatter}, the scatter of the plane-fit parameters for individual hosts is comparable to the scatter of the medians of these properties and to the scatter present among the randomized systems. The sub-halos in the different realisations are not naturally distributed in a way comparable to the observed VPOS. The simulated systems tend to be slightly closer to the extreme VPOS parameters than the randomized ones, but the measured VPOS parameters are more extreme than all median values and than almost all individual realisations (see Table \ref{tab:median} for the median values and extrema for each host).

\subsubsection{Absolute and relative shape}

To be at least as flattened as the VPOS in \textit{absolute} dimension, sub-halo systems must have $r_{\mathrm{per}} \leq r_{\mathrm{per}}^{\mathrm{obs}}$ (Criterion 1 in Table \ref{tab:VPOSresults}). Less than 1\% of the paired or isolated realizations fulfil this criterion. It is apparent from the cumulative distribution of $r_{\mathrm{per}}$\ (Fig. \ref{fig:cumulative}) that sub-halo systems with $r_{\mathrm{per}} \lesssim 35$\,kpc are slightly less frequent among paired than isolated systems. Afterwards, the curve for isolated systems rises more slowly than that of the paired ones, so the average $r_{\mathrm{per}}$\ is slightly smaller for paired than for isolated systems. The cumulative distribution of the randomized systems is similar to the paired one but offset by about 7\,kpc to larger values.

Most systems fulfil criterion 2 ($r_{\mathrm{par}} \geq r_{\mathrm{par}}^{\mathrm{obs}}$) and are thus sufficiently radially extended to be comparable to the VPOS. More narrow planes (small $r_{\mathrm{per}}$) tend to be found for more radially concentrated (small $r_{\mathrm{par}}$) satellite distributions. This can be best seen from the diagonal shape of the contours for the randomized systems and the distribution of the median values for individual hosts in the first panel of Fig. \ref{fig:scatter}.

Criteria 3 and 4 in Table \ref{tab:VPOSresults} use the axis ratios to compare the \textit{relative} shape of the sub-halo systems with that of the MW satellites. The results for $c/a$\ are similar to those for $r_{\mathrm{per}}$, both properties scatter around a linear relation (Fig. \ref{fig:scatter}), and also the cumulative distributions show the same behaviour (Fig. \ref{fig:cumulative}). Again the measured flattening of the VPOS is not naturally found among the sub-halo systems, only 0.2\% of the paired and 1.9\% of the isolated realisations have $(c/a) \leq (c/a)^{\mathrm{obs}}$. The distribution and scatter in $c/a$\ and $b/a$\ (Fig. \ref{fig:scatter}) is similar among paired, isolated, and randomized systems, but the latter tend to have slightly larger $c/a$\ than the simulated sub-halo systems.

\subsubsection{Orbital pole concentration and alignment}

In addition to being similarly flattened, the sub-halo systems also need to have at least as concentrated orbital poles  ($\Delta_{\mathrm{std}} \leq \Delta_{\mathrm{std}}^{\mathrm{obs}}$, criterion 5 in Table \ref{tab:VPOSresults}) which are at least as well aligned with the normal to the best-fit plane ($\theta_{\mathrm{VPOS}} \leq \theta_{\mathrm{VPOS}}^{\mathrm{obs}}$, criterion 6) to be similar to the observed VPOS.

Like for the flattening criteria, paired and isolated systems do not naturally fulfil the $\Delta_{\mathrm{std}}$\ criterion ($\lesssim 1.5$\%, see Table \ref{tab:VPOSresults}). Paired systems are a bit more likely to have the most-concentrated poles ($\Delta_{\mathrm{std}} \lesssim 30^{\circ}$), but in general the isolated systems tend to have slightly smaller $\Delta_{\mathrm{std}}$ (Fig. \ref{fig:cumulative}). Randomized systems have $\approx 5^\circ$\ larger $\Delta_{\mathrm{std}}$\ on average, but again the general behaviour (Fig. \ref{fig:cumulative}) and scatter (Fig. \ref{fig:scatter}) are similar.

The cumulative distributions of $\theta_{\mathrm{VPOS}}$\ for paired and isolated systems (last panel Fig. \ref{fig:cumulative}) are almost identical. There is no indication that the existence of a neighbouring main halo affects the orbital alignment of sub-halos with their preferred plane.
Realisations with $\theta_{\mathrm{VPOS}}^{\mathrm{obs}} = 18.9^{\circ}$\ are almost twice as likely in simulated systems than in the randomized ones (17 to 10\%, see Table \ref{tab:VPOSresults}).

\subsubsection{Combined Criteria}

Only if a plane of sub-halos \textit{simultaneously} meets the different criteria defining the VPOS properties can it be said to reproduce the observed situation. The two essential properties of the VPOS are its narrow extent (measured with $r_{\mathrm{per}}$\ or $c/a$) and the alignment of the orbital poles of the satellites (measured with $\Delta_{\mathrm{std}}$). The last panel in Fig. \ref{fig:scatter} plots $\Delta_{\mathrm{std}}$\ against $r_{\mathrm{per}}$\ and shows that planes in simulated and randomized systems tend to have parameters that are significantly larger than those of the observed MW satellite system. Sub-halo systems that are sufficiently flattened are not sufficiently co-orbiting, while those which co-orbit are not sufficiently flattened. None of the 4800 randomized systems reproduce either of the two combined criteria 8 and 9 in Table \ref{tab:VPOSresults}, setting an upper limit on the fraction of such systems of 0.02\%. Likewise, none of the 2400 systems around isolated hosts fulfils the combined criteria (upper limit of 0.04\%). The sub-halo system with median values in $r_{\mathrm{per}}$\ and $c/a$\ coming closest to the observed ones belongs to the isolated host iRomulus, but none of its realisations is simultaneously sufficiently flattened and has sufficiently concentrated orbital poles.
Among the systems around paired hosts only one out of 2000 realisations (0.05\%) fulfils the combined criteria.

An agreement with the VPOS in two properties is thus extremely rare and requires a finely-tuned obscuring disc orientation: in 1 out of 20 hosts only 1 out of 100 randomly oriented obscuring discs produces a sample of sub-halos that shares two properties with the VPOS. The host of this particular realisation is Oates, whose sub-halo system has a relatively low median $r_{\mathrm{per}}$\ of 42\,kpc, and the lowest median $\Delta_{\mathrm{std}}$\ of $38.7^{\circ}$\ of all paired hosts (see Table \ref{tab:median}). Oates it the fourth-lowest-mass paired halo in ELVIS ('virial' mass of $M_{\mathrm{V}} = 1.2 \times 10^{12}\,M_{\sun}$) and has formed only recently (acquired half of its mass at a redshift of $z = 0.62$) \citep{GarrisonKimmel2014}. Furthermore, the realisation does not fulfil criterion 6 simultaneously, the concentrated orbital poles do not align with the plane normal as closely as observed. 

Even with significantly relaxed criteria ($r_{\mathrm{per}} \leq 1.5 \times r_{\mathrm{per}}^{\mathrm{obs}}$\ and $\Delta_{\mathrm{std}} \leq 1.5 \times \Delta_{\mathrm{std}}^{\mathrm{obs}}$), the fractions of realisations reproducing these simultaneously remain between 0.5\% (paired) and 2\% (isolated).

\subsection{Dependency of satellite system shape on the number of satellites}
\label{sect:shape}

\begin{figure*}
   \centering
   \includegraphics[width=80mm]{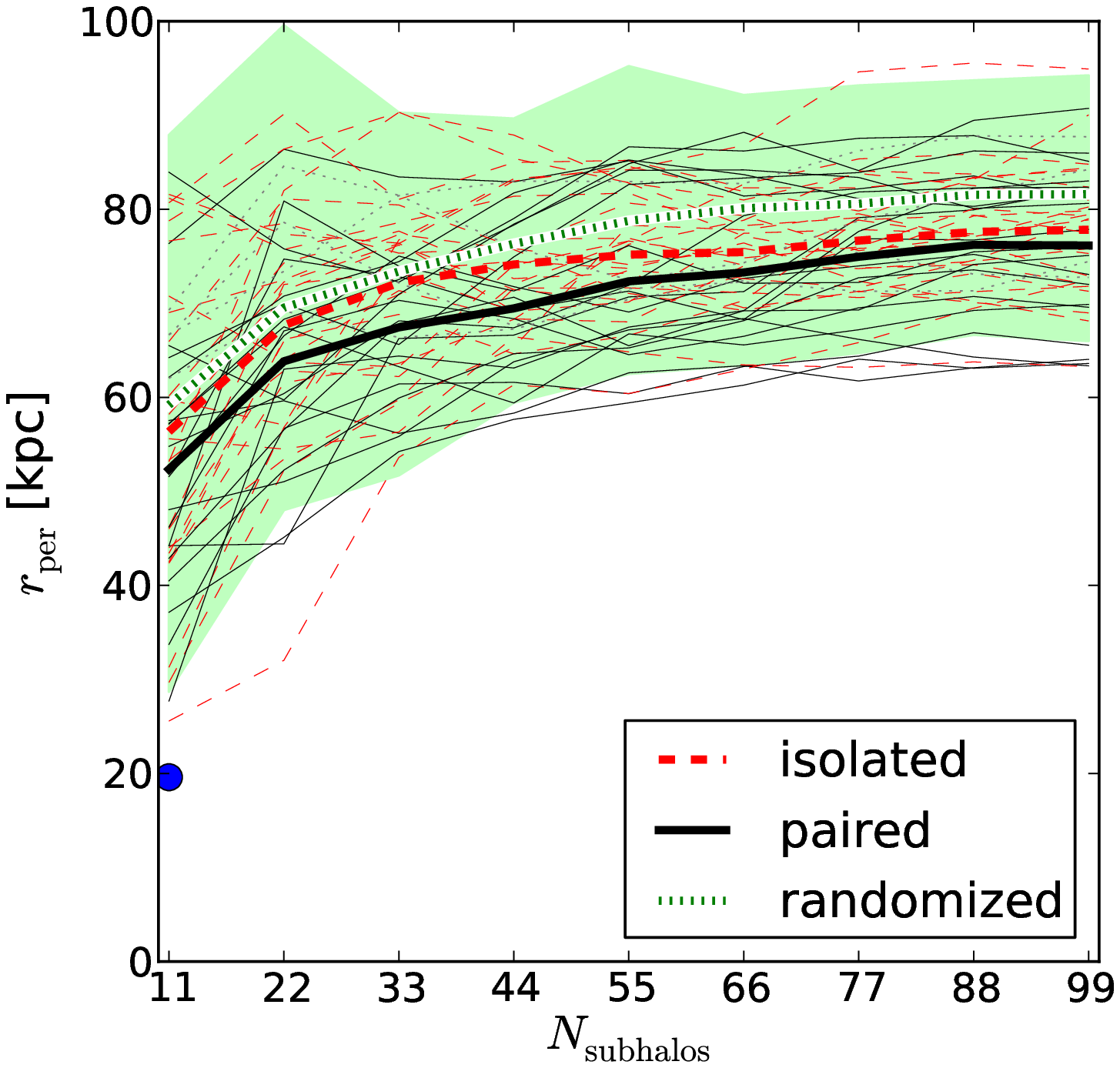}
   \includegraphics[width=80mm]{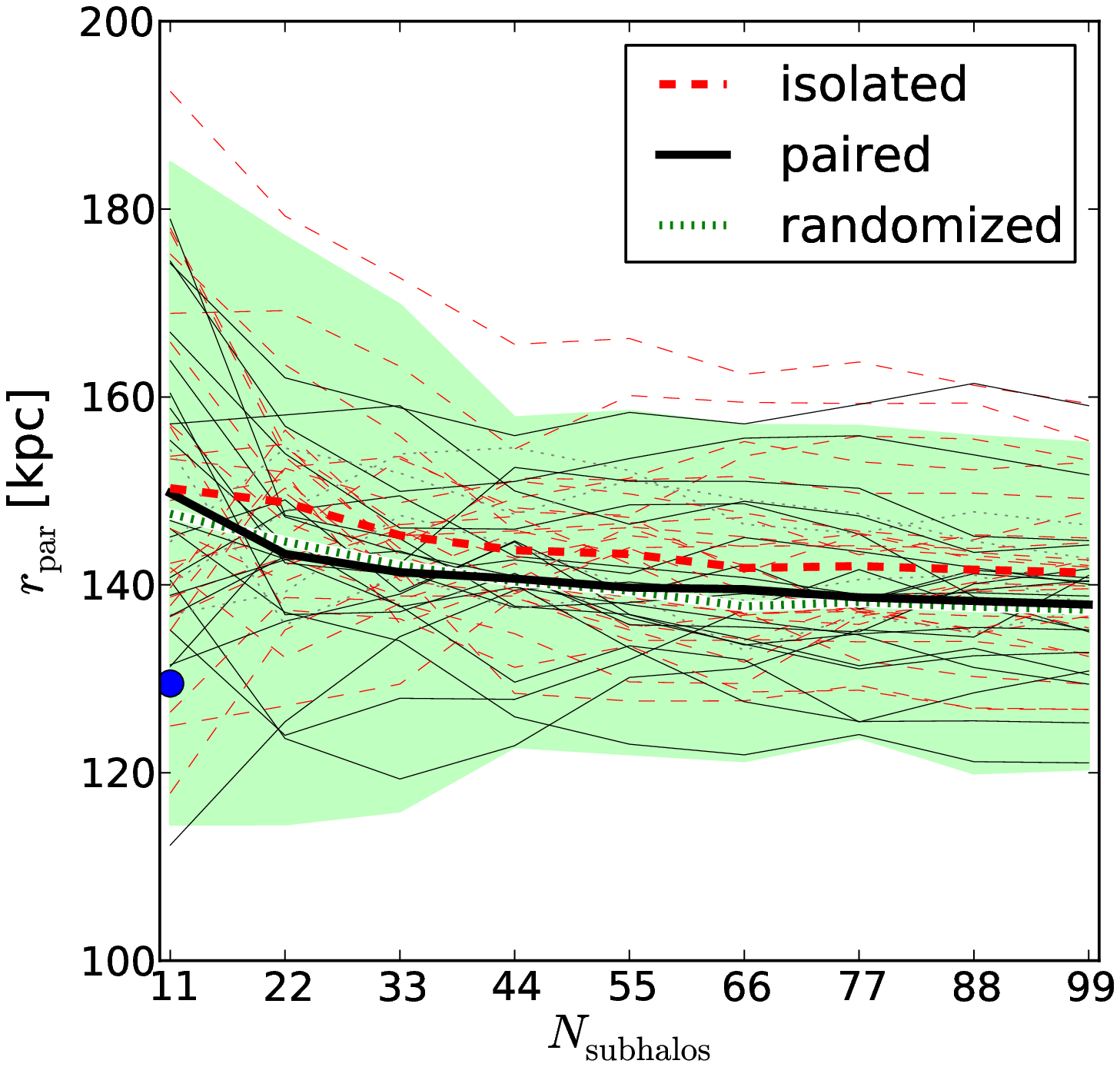}
   \includegraphics[width=80mm]{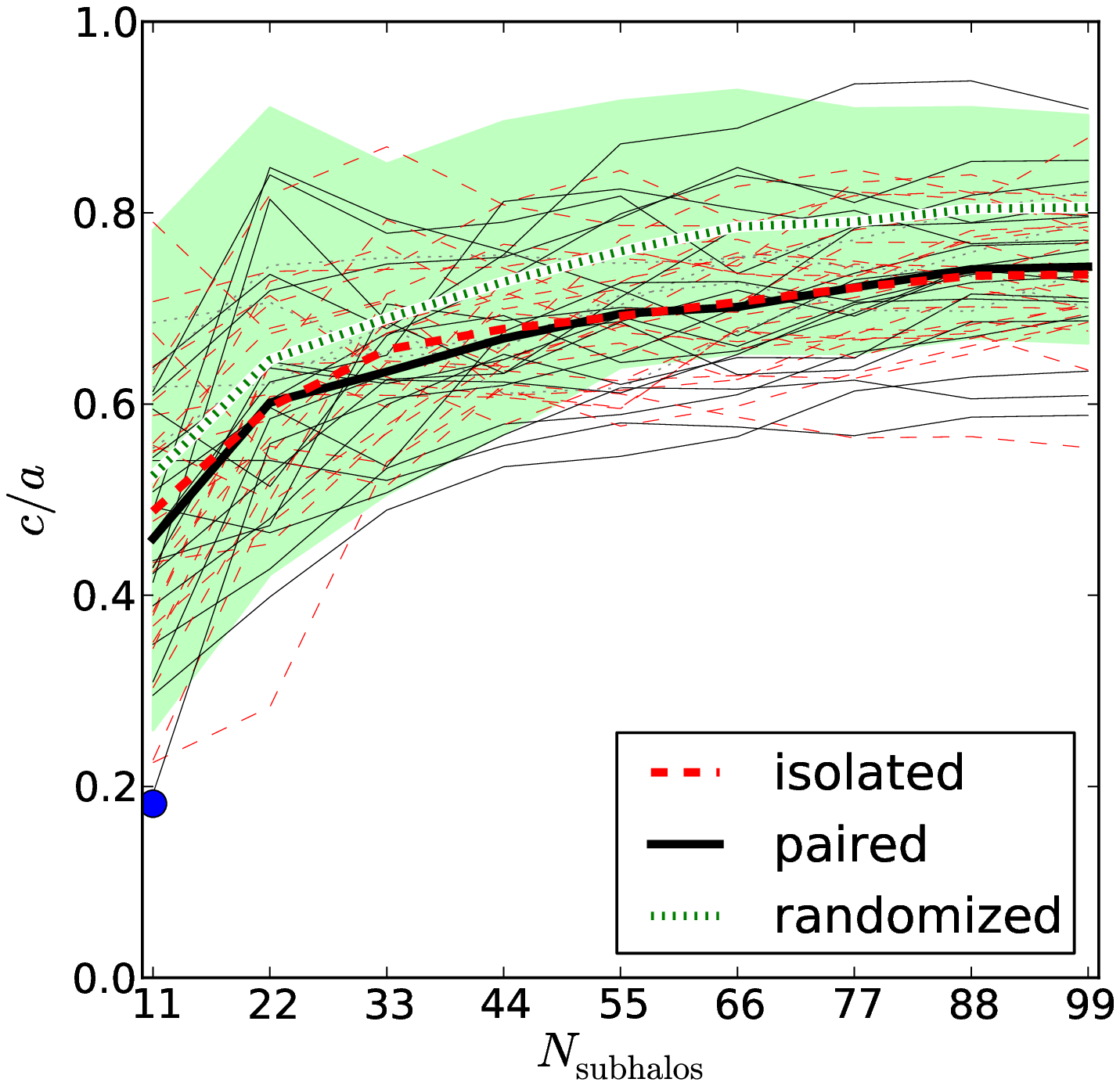}
   \includegraphics[width=80mm]{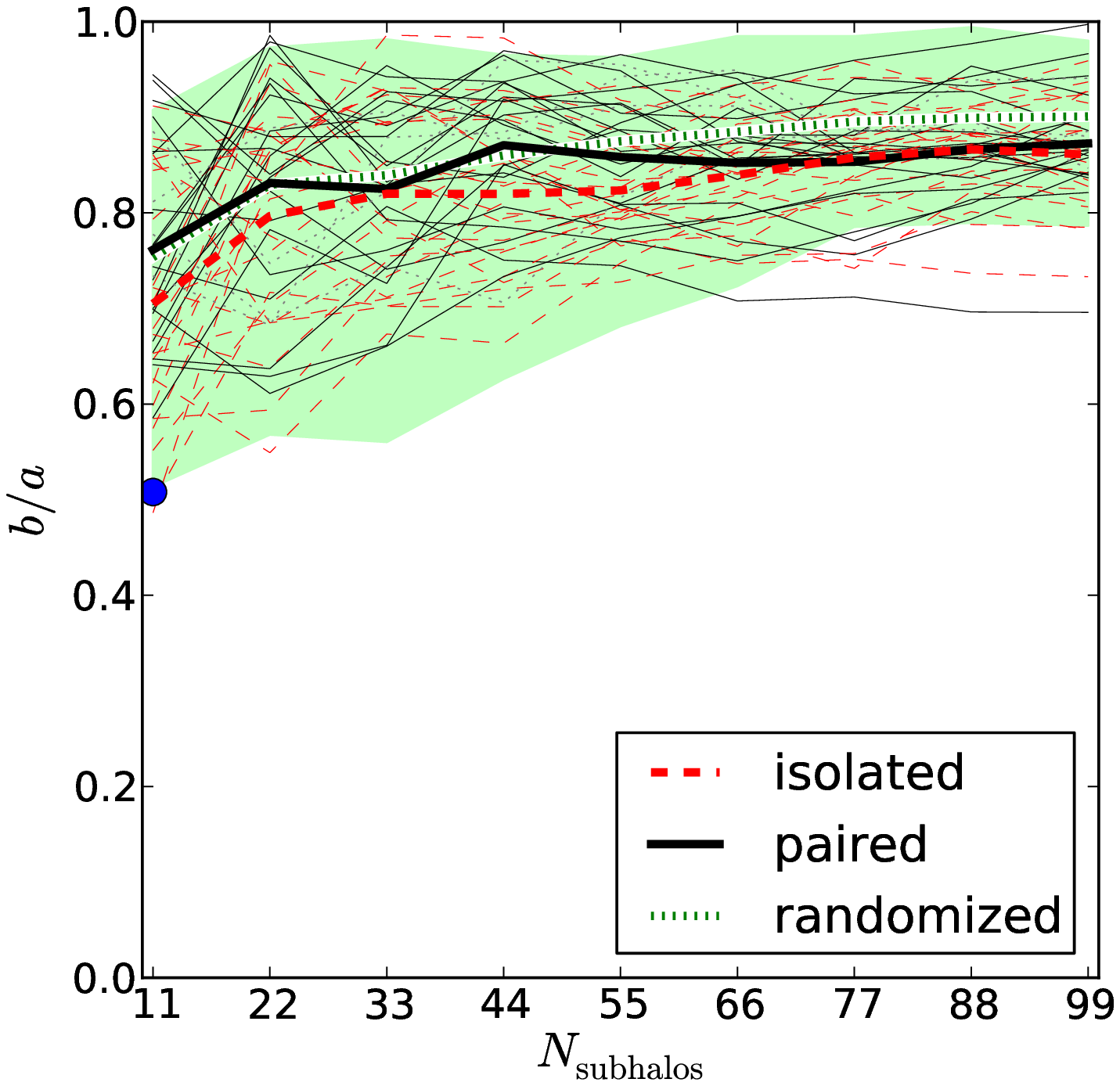}
   \caption{The shape of the distribution of the top $N_{\mathrm{subhalo}}$\ sub-halos ranked by $M_{\mathrm{peak}}$\ in steps of 11. Shown are the RMS height $r_{\mathrm{per}}$, the RMS radius $r_{\mathrm{par}}$, and the short- and intermediate-to-long axis ratios $c/a$\ and $b/a$\ for each paired and isolated host (thin lines) and the averages of these two classes and the randomized systems (thick lines). The shaded area marks the range between the maximum and minimum parameters found for 48 randomized satellite systems. The blue dots indicate the parameters measured for the 11 classical MW satellites in the VPOS. The MW system is highly unusual by three out of four measures.
   }
              \label{fig:shape}
\end{figure*}

We now turn our attention to the dependency of the overall shape of the sub-halo distributions on the number of considered sub-halos. Fig. \ref{fig:shape} plots $r_{\mathrm{per}}$, $r_{\mathrm{par}}$, $c/a$\ and $b/a$\ for the $N_{\mathrm{subhalo}}$ = 11, 22, ..., 99 sub-halos with the largest $M_{\mathrm{peak}}$\ (largest stellar masses according to AM) for each host and the averages for the 20 paired, 24 isolated and 48 randomized systems.

As the number of satellites increases, the average $r_{\mathrm{per}}$\ first rises quickly from about 50--55 to 65--70\,kpc, and then more slowly to about 75\,kpc. The average $r_{\mathrm{per}}$\ of paired hosts remains below that of isolated hosts for all $N_{\mathrm{subhalo}}$, which remains below the average of the randomized systems. Overall the differences are small. The average of the randomized systems is only $\approx 5$\,kpc ($\lesssim 10$\%) larger than that of the paired systems. The difference between the averages for paired and isolated systems becomes smaller for larger $N_{\mathrm{subhalo}}$.

That the average absolute thickness $r_{\mathrm{per}}$\ of paired systems is lower than that of the isolated ones could either indicate that the sub-halos are on average in a more flattened configuration, or that the systems are more radially concentrated on average. The behaviour of the axis ratio $c/a$ hints at the latter explanation. The average $c/a$\ of paired and isolated systems follow the same curve, again rising steeply between $N_{\mathrm{subhalo}} = 11$ and 22 from 0.45 to 0.6 and then approaching a plateau of about 0.7 for large $N_{\mathrm{subhalo}}$\ \citep[see also][]{Wang2013}. The average \textit{relative} thickness of the sub-halo distributions is therefore independent of whether the host is part of a paired group or isolated. For randomized systems $c/a$\ is on average 0.05 to 0.1 larger ($\approx 10$\%), confirming that sub-halo systems are slightly more flattened than isotropic distributions \citep{Zentner2005}.

The average $r_{\mathrm{par}}$\ is largest for small $N_{\mathrm{subhalo}}$, i.e. sub-halos with the largest $M_{\mathrm{peak}}$\ are more radially extended \textit{in the best-fit plane}. However, the effect is minuscule, the average $r_{\mathrm{per}}$\ only changes from 150 to 140\,kpc, and might be due to the decreasing flattening for larger $N_{\mathrm{subhalo}}$\ which causes a larger component of the radial distance to contribute to $r_{\mathrm{per}}$\ instead of $r_{\mathrm{par}}$. The very small difference between paired and isolated systems is not necessarily caused by different environments, but might simply be an effect of low statistics due to the relatively small number of hosts.

The overall behaviour of the randomized and simulated systems is similar in all four parameters, which is also true for the spread in their values. This indicates that an isotropic distribution can be an acceptable zeroth-order approximation for sub-halo systems. The dominating reason why more flattened systems are found for lower $N_{\mathrm{subhalo}}$\ is then probably an effect of the number of sub-halos. An extreme case of only three sub-halos would always result in perfect planes ($r_{\mathrm{per}} = 0$\,kpc and $c/a = 0$). For small $N_{\mathrm{subhalo}}$\ only a few sub-halos are situated at large distances, but these dominate the plane fit. Sub-halos at smaller distances have small offsets from any plane passing close to the center of the host, such that the overall thickness of the distribution tends to be smaller for smaller $N_{\mathrm{subhalo}}$. For larger $N_{\mathrm{subhalo}}$\ more sub-halos will be present at large distances but outside the plane-fit, increasing the measured thickness.


\section{Conclusion}
\label{sect:conclusion}

We have investigated the phase-space distribution of the most-massive sub-halos (ranked by $M_{\mathrm{peak}}$, corresponding to a ranking in stellar mass in AM) around paired and isolated hosts in the ELVIS simulation suite \citep{GarrisonKimmel2014}. If the number of considered sub-halos is small ($N_{\mathrm{subhalo}} \lesssim 20$), the flattening of the sub-halo system strongly depends on and rises for larger $N_{\mathrm{subhalo}}$. This overall behaviour is identical for paired, isolated and randomized (isotropic) systems. The latter have an average flattening offset to slightly larger values, but the scatter among systems of each type is larger than this difference.

We have also compared the phase-space distribution of the 11 top-ranked sub-halos (accounting for obscuration by a galactic disc) with that of the 11 most-massive MW satellites. Paired hosts similar to the LG do not have a higher chance to contain sub-halo distributions which are similarly flattened as the MW satellites in the VPOS. In the analysed simulations, isolated hosts are in fact more likely to have the smallest $r_{\mathrm{per}}$\ and $c/a$, while the corresponding averages are smaller for paired hosts, which might be due to a slightly stronger radial concentration of their sub-halo systems. Paired and isolated systems also show a similar degree of orbital pole alignments.

The low rate of satellite planes that are as strongly correlated as the VPOS found in cosmological simulations such as the Millennium-II \citep{Pawlowski2014b} and the Aquarius simulations \citep{Pawlowski2013b}, is therefore most-likely not affected by ignoring the host halo environments. The absence of VPOS-like structures appears to be a natural feature of dissipationless cosmological simulations. In particular, the VPOS can not satisfactorily be understood as an extreme statistical outlier of the simulated distributions because additional objects align with the structure and more correlated satellite planes have been found in the local Universe \citep[e.g.][]{Kroupa2010,Pawlowski2012a,Ibata2013}.
This emphasizes that the search for an explanation of such structures requires different approaches. Examples for this are the inclusion of gas in cosmological simulations \citep{Khandai2014,Vogelsberger2014} or scenarios which question the association of dwarf galaxies with sub-halos, such as the formation of phase-space correlated populations of tidal dwarf galaxies \citep{Pawlowski2011,Fouquet2012,Hammer2013,Yang2014}.


\acknowledgments
We thank Shea Garrison-Kimmel and the ELVIS collaboration for making their simulations publicly available.

\clearpage

\begin{table}
\small
  \begin{center}
  \caption{\label{tab:VPOSresults}Fractions of sub-halo realisations reproducing the observed VPOS parameters.}
  \begin{tabular}{rlrrrrrr}
  \tableline\tableline
  \# & Criterion \& VPOS parameter & $P_{\mathrm{all}}$ [\%] & $P_{\mathrm{all}}^{\mathrm{rand}}$ [\%] & $P_{\mathrm{pair}}$ [\%] & $P_{\mathrm{pair}}^{\mathrm{rand}}$ [\%] & $P_{\mathrm{isol}}$ [\%] & $P_{\mathrm{isol}}^{\mathrm{rand}}$ [\%] \\
  \tableline
1 & $r_{\mathrm{per}} \leq 19.6$\,kpc & 0.54 & 0.17 & 0.20 & 0.10 & 0.92 & 0.25 \\
2 & 1 \& $r_{\mathrm{par}} \geq 129.5$\,kpc & 0.33 & 0.06 & 0.20 & 0.10 & 0.50 & 0.04 \\
3 & $c/a \leq 0.182$ & 1.04 & 0.25 & 0.20 & 0.15 & 1.92 & 0.38 \\
4 & 3 \& $b/a \geq 0.508$ & 0.77 & 0.23 & 0.20 & 0.15 & 1.38 & 0.33 \\
5 & $\Delta_{\mathrm{std}} \leq 29.3^{\circ}$ & 1.31 & 0.10 & 1.45 & 0.05 & 1.00 & 0.17 \\
6 & $\theta_{\mathrm{VPOS}} \leq 18.9^{\circ}$ & 16.90 & 9.81 & 16.90 & 10.55 & 17.25 & 9.58 \\
7 & 5 \& 6 & 0.60 & 0.04 & 0.70 & 0.05 & 0.62 & 0.04 \\
8 & 1 \& 5 & 0.02 & $< 0.02$ & 0.05 & $< 0.05$ & $< 0.04$ & $< 0.04$ \\
9 & 3 \& 5 & 0.02 & $< 0.02$ & 0.05 & $< 0.05$ & $< 0.04$ & $< 0.04$ \\
  \tableline
 & $r_{\mathrm{per}} \leq 1.5\,r_{\mathrm{per}}^{\mathrm{obs}}$ \& $\Delta_{\mathrm{std}} \leq 1.5\,\Delta_{\mathrm{std}}^{\mathrm{obs}}$ & 1.33 & 0.23 & 0.5 & 0.1 & 2.25 & 0.38 \\
  \tableline
\end{tabular}
 \tablecomments{The criteria required to reproduce the VPOS plane-fit parameters (column 2, numbered in column 1) for all 48, the 20 paired and the 24 isolated hosts (columns 3, 5 and 7) as well as the corresponding randomized realisations (columns 4, 6 and 8). Each host contributes 100 realisations with randomly oriented obscuring discs.}
 \end{center}
\end{table}

\clearpage

\begin{deluxetable}{lcccccc}
\tablecaption{Median plane parameters per host\label{tab:median}}
\tablewidth{0pt}
\tablehead{
\colhead{Host} & 
\colhead{$r_{\mathrm{per}}$~[kpc]}& 
\colhead{$r_{\mathrm{par}}$~[kpc]}& 
\colhead{$c/a$}& 
\colhead{$b/a$}& 
\colhead{$\Delta_{\mathrm{std}}~[^{\circ}]$}& 
\colhead{$\theta_{\mathrm{VPOS}}~[^{\circ}]$} 
}
\startdata
Hera& $59.8_{\mathrm{-26.3}}^{\mathrm{+25.3}} $& $172.7_{\mathrm{-28.1}}^{\mathrm{+23.4}} $& $ 0.51_{\mathrm{-0.25}}^{\mathrm{+0.19}} $& $ 0.68_{\mathrm{-0.22}}^{\mathrm{+0.23}} $& $46.4_{\mathrm{-20.5}}^{\mathrm{+16.2}} $& $56.3_{\mathrm{-51.4}}^{\mathrm{+31.8}} $ \\ 
Zeus& $50.5_{\mathrm{-15.0}}^{\mathrm{+12.9}} $& $140.9_{\mathrm{-25.9}}^{\mathrm{+15.6}} $& $ 0.48_{\mathrm{-0.17}}^{\mathrm{+0.17}} $& $ 0.80_{\mathrm{-0.22}}^{\mathrm{+0.19}} $& $53.8_{\mathrm{-14.6}}^{\mathrm{+13.3}} $& $73.5_{\mathrm{-48.1}}^{\mathrm{+15.9}} $ \\ 
Scylla& $33.7_{\mathrm{-16.0}}^{\mathrm{+15.7}} $& $145.5_{\mathrm{-39.1}}^{\mathrm{+17.9}} $& $ 0.31_{\mathrm{-0.16}}^{\mathrm{+0.15}} $& $ 0.83_{\mathrm{-0.33}}^{\mathrm{+0.15}} $& $58.8_{\mathrm{-17.3}}^{\mathrm{+20.0}} $& $60.4_{\mathrm{-55.5}}^{\mathrm{+21.0}} $ \\ 
Charybdis& $54.6_{\mathrm{-12.8}}^{\mathrm{+14.1}} $& $117.3_{\mathrm{-8.9}}^{\mathrm{+35.8}} $& $ 0.58_{\mathrm{-0.14}}^{\mathrm{+0.33}} $& $ 0.75_{\mathrm{-0.13}}^{\mathrm{+0.24}} $& $57.5_{\mathrm{-6.8}}^{\mathrm{+9.1}} $& $60.2_{\mathrm{-42.8}}^{\mathrm{+28.9}} $ \\ 
Romulus& $76.7_{\mathrm{-27.1}}^{\mathrm{+18.7}} $& $160.6_{\mathrm{-13.3}}^{\mathrm{+9.9}} $& $ 0.62_{\mathrm{-0.25}}^{\mathrm{+0.24}} $& $ 0.73_{\mathrm{-0.13}}^{\mathrm{+0.25}} $& $52.7_{\mathrm{-8.7}}^{\mathrm{+13.2}} $& $63.9_{\mathrm{-40.5}}^{\mathrm{+24.6}} $ \\ 
Remus& $49.0_{\mathrm{-25.0}}^{\mathrm{+12.4}} $& $145.8_{\mathrm{-28.0}}^{\mathrm{+18.5}} $& $ 0.47_{\mathrm{-0.26}}^{\mathrm{+0.18}} $& $ 0.81_{\mathrm{-0.44}}^{\mathrm{+0.16}} $& $55.4_{\mathrm{-12.1}}^{\mathrm{+16.4}} $& $43.8_{\mathrm{-42.9}}^{\mathrm{+41.7}} $ \\ 
Orion& $40.3_{\mathrm{-7.4}}^{\mathrm{+19.1}} $& $170.3_{\mathrm{-31.4}}^{\mathrm{+14.5}} $& $ 0.29_{\mathrm{-0.05}}^{\mathrm{+0.25}} $& $ 0.64_{\mathrm{-0.26}}^{\mathrm{+0.25}} $& $52.8_{\mathrm{-20.9}}^{\mathrm{+12.2}} $& $38.9_{\mathrm{-28.5}}^{\mathrm{+45.7}} $ \\ 
Taurus& $55.0_{\mathrm{-30.4}}^{\mathrm{+13.0}} $& $141.2_{\mathrm{-23.4}}^{\mathrm{+15.9}} $& $ 0.51_{\mathrm{-0.26}}^{\mathrm{+0.17}} $& $ 0.80_{\mathrm{-0.25}}^{\mathrm{+0.18}} $& $60.6_{\mathrm{-14.0}}^{\mathrm{+8.5}} $& $38.2_{\mathrm{-35.0}}^{\mathrm{+47.8}} $ \\ 
Kek& $50.4_{\mathrm{-25.7}}^{\mathrm{+24.4}} $& $150.2_{\mathrm{-28.6}}^{\mathrm{+10.2}} $& $ 0.39_{\mathrm{-0.20}}^{\mathrm{+0.27}} $& $ 0.65_{\mathrm{-0.32}}^{\mathrm{+0.28}} $& $59.3_{\mathrm{-23.1}}^{\mathrm{+9.0}} $& $25.2_{\mathrm{-21.0}}^{\mathrm{+61.6}} $ \\ 
Kauket& $45.8_{\mathrm{-18.1}}^{\mathrm{+26.1}} $& $171.2_{\mathrm{-35.5}}^{\mathrm{+16.7}} $& $ 0.34_{\mathrm{-0.15}}^{\mathrm{+0.41}} $& $ 0.71_{\mathrm{-0.28}}^{\mathrm{+0.28}} $& $41.7_{\mathrm{-5.5}}^{\mathrm{+13.8}} $& $35.7_{\mathrm{-24.6}}^{\mathrm{+53.2}} $ \\ 
Hamilton& $51.6_{\mathrm{-21.9}}^{\mathrm{+31.2}} $& $166.9_{\mathrm{-26.9}}^{\mathrm{+11.9}} $& $ 0.45_{\mathrm{-0.20}}^{\mathrm{+0.26}} $& $ 0.86_{\mathrm{-0.26}}^{\mathrm{+0.11}} $& $41.6_{\mathrm{-9.4}}^{\mathrm{+14.1}} $& $78.6_{\mathrm{-55.0}}^{\mathrm{+11.2}} $ \\ 
Burr& $37.1_{\mathrm{-11.5}}^{\mathrm{+6.6}} $& $137.6_{\mathrm{-15.8}}^{\mathrm{+18.0}} $& $ 0.33_{\mathrm{-0.10}}^{\mathrm{+0.10}} $& $ 0.67_{\mathrm{-0.12}}^{\mathrm{+0.22}} $& $52.5_{\mathrm{-17.7}}^{\mathrm{+18.1}} $& $12.0_{\mathrm{-10.2}}^{\mathrm{+28.7}} $ \\ 
Lincoln& $60.1_{\mathrm{-32.5}}^{\mathrm{+17.1}} $& $150.9_{\mathrm{-24.4}}^{\mathrm{+7.4}} $& $ 0.46_{\mathrm{-0.26}}^{\mathrm{+0.32}} $& $ 0.59_{\mathrm{-0.14}}^{\mathrm{+0.35}} $& $54.2_{\mathrm{-14.5}}^{\mathrm{+10.4}} $& $40.5_{\mathrm{-38.4}}^{\mathrm{+46.7}} $ \\ 
Douglas& $43.0_{\mathrm{-16.2}}^{\mathrm{+14.6}} $& $136.7_{\mathrm{-36.8}}^{\mathrm{+19.8}} $& $ 0.40_{\mathrm{-0.13}}^{\mathrm{+0.13}} $& $ 0.75_{\mathrm{-0.24}}^{\mathrm{+0.18}} $& $51.7_{\mathrm{-6.0}}^{\mathrm{+15.0}} $& $34.9_{\mathrm{-34.3}}^{\mathrm{+52.6}} $ \\ 
Serana\tablenotemark{a}& $65.7_{\mathrm{-23.4}}^{\mathrm{+17.5}} $& $148.1_{\mathrm{-27.3}}^{\mathrm{+29.6}} $& $ 0.57_{\mathrm{-0.23}}^{\mathrm{+0.22}} $& $ 0.74_{\mathrm{-0.21}}^{\mathrm{+0.18}} $& $53.8_{\mathrm{-17.7}}^{\mathrm{+14.2}} $& $61.6_{\mathrm{-45.7}}^{\mathrm{+27.8}} $ \\ 
Venus\tablenotemark{a}& $55.4_{\mathrm{-28.3}}^{\mathrm{+18.1}} $& $148.9_{\mathrm{-24.9}}^{\mathrm{+14.5}} $& $ 0.49_{\mathrm{-0.27}}^{\mathrm{+0.23}} $& $ 0.71_{\mathrm{-0.25}}^{\mathrm{+0.26}} $& $53.2_{\mathrm{-7.3}}^{\mathrm{+12.0}} $& $51.6_{\mathrm{-40.8}}^{\mathrm{+34.2}} $ \\ 
Sonny& $55.1_{\mathrm{-19.6}}^{\mathrm{+15.2}} $& $138.0_{\mathrm{-25.8}}^{\mathrm{+14.5}} $& $ 0.52_{\mathrm{-0.20}}^{\mathrm{+0.24}} $& $ 0.70_{\mathrm{-0.11}}^{\mathrm{+0.29}} $& $58.8_{\mathrm{-9.7}}^{\mathrm{+12.9}} $& $46.8_{\mathrm{-44.1}}^{\mathrm{+42.0}} $ \\ 
Cher& $71.5_{\mathrm{-24.8}}^{\mathrm{+26.3}} $& $158.8_{\mathrm{-18.3}}^{\mathrm{+13.1}} $& $ 0.60_{\mathrm{-0.21}}^{\mathrm{+0.24}} $& $ 0.80_{\mathrm{-0.15}}^{\mathrm{+0.11}} $& $45.9_{\mathrm{-8.1}}^{\mathrm{+18.9}} $& $34.3_{\mathrm{-33.0}}^{\mathrm{+49.0}} $ \\ 
Hall& $51.9_{\mathrm{-31.1}}^{\mathrm{+22.9}} $& $134.6_{\mathrm{-24.3}}^{\mathrm{+22.6}} $& $ 0.49_{\mathrm{-0.28}}^{\mathrm{+0.26}} $& $ 0.77_{\mathrm{-0.25}}^{\mathrm{+0.16}} $& $56.3_{\mathrm{-8.6}}^{\mathrm{+6.6}} $& $53.6_{\mathrm{-47.7}}^{\mathrm{+35.9}} $ \\ 
Oates& $42.0_{\mathrm{-23.3}}^{\mathrm{+9.5}} $& $133.0_{\mathrm{-27.2}}^{\mathrm{+12.7}} $& $ 0.41_{\mathrm{-0.23}}^{\mathrm{+0.15}} $& $ 0.82_{\mathrm{-0.34}}^{\mathrm{+0.16}} $& $38.7_{\mathrm{-15.9}}^{\mathrm{+18.6}} $& $17.7_{\mathrm{-11.4}}^{\mathrm{+69.5}} $ \\ 
Thelma& $50.5_{\mathrm{-26.0}}^{\mathrm{+12.9}} $& $146.8_{\mathrm{-31.4}}^{\mathrm{+13.2}} $& $ 0.44_{\mathrm{-0.24}}^{\mathrm{+0.18}} $& $ 0.76_{\mathrm{-0.33}}^{\mathrm{+0.21}} $& $58.5_{\mathrm{-18.3}}^{\mathrm{+12.6}} $& $35.9_{\mathrm{-20.2}}^{\mathrm{+52.0}} $ \\ 
Louise& $52.9_{\mathrm{-17.6}}^{\mathrm{+11.1}} $& $152.2_{\mathrm{-31.2}}^{\mathrm{+15.9}} $& $ 0.47_{\mathrm{-0.16}}^{\mathrm{+0.20}} $& $ 0.76_{\mathrm{-0.13}}^{\mathrm{+0.22}} $& $54.3_{\mathrm{-6.8}}^{\mathrm{+10.2}} $& $35.6_{\mathrm{-25.4}}^{\mathrm{+53.1}} $ \\ 
Siegfried\tablenotemark{a}& $60.8_{\mathrm{-21.7}}^{\mathrm{+7.4}} $& $147.2_{\mathrm{-23.8}}^{\mathrm{+34.0}} $& $ 0.54_{\mathrm{-0.29}}^{\mathrm{+0.21}} $& $ 0.72_{\mathrm{-0.35}}^{\mathrm{+0.18}} $& $62.7_{\mathrm{-14.6}}^{\mathrm{+7.7}} $& $70.3_{\mathrm{-60.1}}^{\mathrm{+19.6}} $ \\ 
Roy\tablenotemark{a}& $55.9_{\mathrm{-25.1}}^{\mathrm{+9.2}} $& $145.9_{\mathrm{-29.2}}^{\mathrm{+12.5}} $& $ 0.53_{\mathrm{-0.25}}^{\mathrm{+0.16}} $& $ 0.73_{\mathrm{-0.23}}^{\mathrm{+0.22}} $& $36.7_{\mathrm{-8.2}}^{\mathrm{+17.5}} $& $23.9_{\mathrm{-19.4}}^{\mathrm{+59.4}} $ \\ 
iHera& $43.7_{\mathrm{-13.1}}^{\mathrm{+27.0}} $& $173.0_{\mathrm{-29.7}}^{\mathrm{+23.4}} $& $ 0.36_{\mathrm{-0.15}}^{\mathrm{+0.22}} $& $ 0.59_{\mathrm{-0.20}}^{\mathrm{+0.26}} $& $35.6_{\mathrm{-7.1}}^{\mathrm{+22.4}} $& $36.9_{\mathrm{-17.6}}^{\mathrm{+46.2}} $ \\ 
iZeus& $58.7_{\mathrm{-17.1}}^{\mathrm{+9.7}} $& $144.5_{\mathrm{-20.3}}^{\mathrm{+18.3}} $& $ 0.50_{\mathrm{-0.17}}^{\mathrm{+0.19}} $& $ 0.71_{\mathrm{-0.30}}^{\mathrm{+0.16}} $& $47.0_{\mathrm{-6.2}}^{\mathrm{+8.7}} $& $57.7_{\mathrm{-56.1}}^{\mathrm{+31.2}} $ \\ 
iScylla& $48.8_{\mathrm{-14.9}}^{\mathrm{+25.6}} $& $141.8_{\mathrm{-16.6}}^{\mathrm{+17.3}} $& $ 0.43_{\mathrm{-0.15}}^{\mathrm{+0.23}} $& $ 0.67_{\mathrm{-0.14}}^{\mathrm{+0.31}} $& $37.7_{\mathrm{-10.2}}^{\mathrm{+13.1}} $& $21.3_{\mathrm{-19.5}}^{\mathrm{+66.4}} $ \\ 
iCharybdis& $63.0_{\mathrm{-12.7}}^{\mathrm{+12.2}} $& $162.1_{\mathrm{-32.6}}^{\mathrm{+20.1}} $& $ 0.51_{\mathrm{-0.13}}^{\mathrm{+0.19}} $& $ 0.73_{\mathrm{-0.17}}^{\mathrm{+0.26}} $& $45.2_{\mathrm{-9.2}}^{\mathrm{+14.5}} $& $51.2_{\mathrm{-48.0}}^{\mathrm{+37.7}} $ \\ 
iRomulus& $26.8_{\mathrm{-10.5}}^{\mathrm{+17.7}} $& $163.6_{\mathrm{-35.3}}^{\mathrm{+5.2}} $& $ 0.22_{\mathrm{-0.07}}^{\mathrm{+0.19}} $& $ 0.80_{\mathrm{-0.31}}^{\mathrm{+0.15}} $& $51.0_{\mathrm{-26.6}}^{\mathrm{+14.8}} $& $33.9_{\mathrm{-33.4}}^{\mathrm{+29.9}} $ \\ 
iRemus& $56.7_{\mathrm{-34.4}}^{\mathrm{+21.7}} $& $153.7_{\mathrm{-19.3}}^{\mathrm{+19.5}} $& $ 0.44_{\mathrm{-0.27}}^{\mathrm{+0.19}} $& $ 0.56_{\mathrm{-0.18}}^{\mathrm{+0.33}} $& $54.7_{\mathrm{-22.1}}^{\mathrm{+14.8}} $& $21.9_{\mathrm{-20.2}}^{\mathrm{+65.2}} $ \\ 
iOrion& $50.8_{\mathrm{-26.4}}^{\mathrm{+10.6}} $& $136.8_{\mathrm{-21.6}}^{\mathrm{+18.3}} $& $ 0.47_{\mathrm{-0.25}}^{\mathrm{+0.16}} $& $ 0.63_{\mathrm{-0.21}}^{\mathrm{+0.18}} $& $62.3_{\mathrm{-32.0}}^{\mathrm{+11.3}} $& $72.1_{\mathrm{-39.9}}^{\mathrm{+17.5}} $ \\ 
iTaurus& $67.7_{\mathrm{-32.3}}^{\mathrm{+18.4}} $& $153.4_{\mathrm{-16.4}}^{\mathrm{+19.2}} $& $ 0.54_{\mathrm{-0.25}}^{\mathrm{+0.20}} $& $ 0.68_{\mathrm{-0.09}}^{\mathrm{+0.26}} $& $64.5_{\mathrm{-17.3}}^{\mathrm{+6.0}} $& $62.7_{\mathrm{-51.0}}^{\mathrm{+27.1}} $ \\ 
iKek& $64.2_{\mathrm{-28.7}}^{\mathrm{+17.4}} $& $135.3_{\mathrm{-14.1}}^{\mathrm{+18.9}} $& $ 0.62_{\mathrm{-0.31}}^{\mathrm{+0.18}} $& $ 0.84_{\mathrm{-0.36}}^{\mathrm{+0.13}} $& $51.2_{\mathrm{-11.8}}^{\mathrm{+10.5}} $& $50.4_{\mathrm{-38.7}}^{\mathrm{+38.9}} $ \\ 
iKauket& $55.5_{\mathrm{-20.3}}^{\mathrm{+5.8}} $& $137.5_{\mathrm{-32.5}}^{\mathrm{+14.4}} $& $ 0.65_{\mathrm{-0.30}}^{\mathrm{+0.10}} $& $ 0.81_{\mathrm{-0.37}}^{\mathrm{+0.15}} $& $54.7_{\mathrm{-16.8}}^{\mathrm{+16.1}} $& $24.8_{\mathrm{-21.5}}^{\mathrm{+65.0}} $ \\ 
iHamilton& $69.4_{\mathrm{-29.2}}^{\mathrm{+24.8}} $& $165.0_{\mathrm{-19.0}}^{\mathrm{+31.0}} $& $ 0.54_{\mathrm{-0.24}}^{\mathrm{+0.23}} $& $ 0.77_{\mathrm{-0.30}}^{\mathrm{+0.19}} $& $61.0_{\mathrm{-17.6}}^{\mathrm{+12.4}} $& $49.3_{\mathrm{-43.2}}^{\mathrm{+38.4}} $ \\ 
iBurr& $42.4_{\mathrm{-22.8}}^{\mathrm{+16.6}} $& $154.6_{\mathrm{-29.1}}^{\mathrm{+12.3}} $& $ 0.38_{\mathrm{-0.19}}^{\mathrm{+0.22}} $& $ 0.72_{\mathrm{-0.32}}^{\mathrm{+0.14}} $& $52.4_{\mathrm{-18.9}}^{\mathrm{+17.7}} $& $18.8_{\mathrm{-13.0}}^{\mathrm{+58.1}} $ \\ 
iLincoln& $44.1_{\mathrm{-25.2}}^{\mathrm{+18.4}} $& $171.5_{\mathrm{-53.4}}^{\mathrm{+18.8}} $& $ 0.33_{\mathrm{-0.20}}^{\mathrm{+0.25}} $& $ 0.81_{\mathrm{-0.27}}^{\mathrm{+0.13}} $& $59.9_{\mathrm{-11.7}}^{\mathrm{+10.6}} $& $47.7_{\mathrm{-36.2}}^{\mathrm{+40.7}} $ \\ 
iDouglas& $62.7_{\mathrm{-34.2}}^{\mathrm{+15.1}} $& $142.7_{\mathrm{-22.4}}^{\mathrm{+24.0}} $& $ 0.55_{\mathrm{-0.27}}^{\mathrm{+0.18}} $& $ 0.70_{\mathrm{-0.20}}^{\mathrm{+0.24}} $& $51.9_{\mathrm{-10.5}}^{\mathrm{+13.8}} $& $75.7_{\mathrm{-68.0}}^{\mathrm{+14.0}} $ \\ 
iSerana& $43.7_{\mathrm{-20.0}}^{\mathrm{+25.1}} $& $143.0_{\mathrm{-15.0}}^{\mathrm{+37.9}} $& $ 0.43_{\mathrm{-0.20}}^{\mathrm{+0.25}} $& $ 0.83_{\mathrm{-0.28}}^{\mathrm{+0.11}} $& $51.1_{\mathrm{-15.3}}^{\mathrm{+15.9}} $& $51.8_{\mathrm{-43.0}}^{\mathrm{+38.0}} $ \\ 
iVenus& $67.5_{\mathrm{-24.6}}^{\mathrm{+9.9}} $& $147.6_{\mathrm{-18.5}}^{\mathrm{+18.2}} $& $ 0.60_{\mathrm{-0.22}}^{\mathrm{+0.20}} $& $ 0.81_{\mathrm{-0.22}}^{\mathrm{+0.17}} $& $48.2_{\mathrm{-9.5}}^{\mathrm{+7.2}} $& $65.1_{\mathrm{-33.5}}^{\mathrm{+24.7}} $ \\ 
iSonny& $31.3_{\mathrm{-20.8}}^{\mathrm{+36.6}} $& $117.8_{\mathrm{-26.5}}^{\mathrm{+27.4}} $& $ 0.30_{\mathrm{-0.22}}^{\mathrm{+0.39}} $& $ 0.50_{\mathrm{-0.18}}^{\mathrm{+0.41}} $& $60.6_{\mathrm{-15.3}}^{\mathrm{+10.5}} $& $28.8_{\mathrm{-24.9}}^{\mathrm{+57.4}} $ \\ 
iCher& $54.4_{\mathrm{-13.7}}^{\mathrm{+28.0}} $& $185.5_{\mathrm{-30.9}}^{\mathrm{+21.1}} $& $ 0.37_{\mathrm{-0.10}}^{\mathrm{+0.33}} $& $ 0.58_{\mathrm{-0.22}}^{\mathrm{+0.33}} $& $61.3_{\mathrm{-24.9}}^{\mathrm{+12.3}} $& $21.6_{\mathrm{-17.8}}^{\mathrm{+57.4}} $ \\ 
iHall& $53.8_{\mathrm{-21.8}}^{\mathrm{+29.9}} $& $146.0_{\mathrm{-23.1}}^{\mathrm{+21.6}} $& $ 0.50_{\mathrm{-0.22}}^{\mathrm{+0.22}} $& $ 0.75_{\mathrm{-0.22}}^{\mathrm{+0.23}} $& $52.2_{\mathrm{-14.0}}^{\mathrm{+13.6}} $& $29.6_{\mathrm{-26.5}}^{\mathrm{+59.2}} $ \\ 
iOates& $46.7_{\mathrm{-15.1}}^{\mathrm{+11.8}} $& $128.9_{\mathrm{-23.0}}^{\mathrm{+17.6}} $& $ 0.44_{\mathrm{-0.15}}^{\mathrm{+0.17}} $& $ 0.68_{\mathrm{-0.20}}^{\mathrm{+0.26}} $& $49.6_{\mathrm{-11.2}}^{\mathrm{+17.4}} $& $61.6_{\mathrm{-54.8}}^{\mathrm{+28.3}} $ \\ 
iThelma& $56.9_{\mathrm{-26.7}}^{\mathrm{+19.5}} $& $126.7_{\mathrm{-12.0}}^{\mathrm{+29.0}} $& $ 0.58_{\mathrm{-0.26}}^{\mathrm{+0.21}} $& $ 0.79_{\mathrm{-0.19}}^{\mathrm{+0.17}} $& $60.1_{\mathrm{-18.0}}^{\mathrm{+11.1}} $& $39.4_{\mathrm{-32.8}}^{\mathrm{+49.5}} $ \\ 
iLouise& $70.4_{\mathrm{-49.5}}^{\mathrm{+18.1}} $& $173.9_{\mathrm{-23.5}}^{\mathrm{+12.1}} $& $ 0.50_{\mathrm{-0.34}}^{\mathrm{+0.32}} $& $ 0.66_{\mathrm{-0.23}}^{\mathrm{+0.25}} $& $42.2_{\mathrm{-20.3}}^{\mathrm{+16.4}} $& $29.4_{\mathrm{-27.6}}^{\mathrm{+58.8}} $ \\ 
iRoy& $34.2_{\mathrm{-10.8}}^{\mathrm{+5.5}} $& $159.1_{\mathrm{-27.0}}^{\mathrm{+14.4}} $& $ 0.27_{\mathrm{-0.10}}^{\mathrm{+0.13}} $& $ 0.70_{\mathrm{-0.25}}^{\mathrm{+0.25}} $& $49.2_{\mathrm{-22.6}}^{\mathrm{+13.5}} $& $37.2_{\mathrm{-19.6}}^{\mathrm{+35.2}} $ \\ 
iSiegfried& $73.5_{\mathrm{-27.3}}^{\mathrm{+16.9}} $& $157.1_{\mathrm{-18.9}}^{\mathrm{+14.6}} $& $ 0.60_{\mathrm{-0.22}}^{\mathrm{+0.19}} $& $ 0.72_{\mathrm{-0.10}}^{\mathrm{+0.26}} $& $55.3_{\mathrm{-14.9}}^{\mathrm{+11.0}} $& $45.7_{\mathrm{-40.2}}^{\mathrm{+43.5}} $ \\ 
\enddata
\tablecomments{Median parameters of planes fitted to the top 11 sub-halos, determined from 100 realisations with randomly oriented obscuring discs. The uncertainties indicate the maximum and minimum value reached for each system. Names of isolated hosts start with 'i'.}
\tablenotetext{a}{Excluded from paired sample due to third massive halo at $\approx 1$\,Mpc distance.}
\end{deluxetable}


\begin{thebibliography}{}

\bibitem[Bahl 
\& Baumgardt(2014)]{Bahl2014} Bahl, H., \& Baumgardt, H.\ 2014, \mnras, 438, 2916 

\bibitem[{{Behroozi} {et~al.}(2013){Behroozi}, {Wechsler}, \&
  {Conroy}}]{Behroozi2013}
{Behroozi}, P.~S., {Wechsler}, R.~H., \& {Conroy}, C. 2013, \apj, 770, 57

\bibitem[{{Deason} {et~al.}(2011){Deason}, {McCarthy}, {Font}, {Evans},
  {Frenk}, {Belokurov}, {Libeskind}, {Crain}, \& {Theuns}}]{Deason2011}
{Deason}, A.~J., {McCarthy}, I.~G., {Font}, A.~S., {et~al.} 2011, \mnras, 415,
  2607

\bibitem[{{D'Onghia} \& {Lake}(2008)}]{DOnghia2008}
{D'Onghia}, E. \& {Lake}, G. 2008, \apjl, 686, L61

\bibitem[{{Fouquet} {et~al.}(2012){Fouquet}, {Hammer}, {Yang}, {Puech}, \&
  {Flores}}]{Fouquet2012}
{Fouquet}, S., {Hammer}, F., {Yang}, Y., {Puech}, M., \& {Flores}, H. 2012,
  \mnras, 427, 1769

\bibitem[{{Garrison-Kimmel} {et~al.}(2014){Garrison-Kimmel}, {Boylan-Kolchin},
  {Bullock}, \& {Lee}}]{GarrisonKimmel2014}
{Garrison-Kimmel}, S., {Boylan-Kolchin}, M., {Bullock}, J.~S., \& {Lee}, K.
  2014, \mnras, 438, 2578

\bibitem[{{Hammer} {et~al.}(2013){Hammer}, {Yang}, {Fouquet}, {Pawlowski},
  {Kroupa}, {Puech}, {Flores}, \& {Wang}}]{Hammer2013}
{Hammer}, F., {Yang}, Y., {Fouquet}, S., {et~al.} 2013, \mnras, 431, 3543

\bibitem[{{Ibata} {et~al.}(2014){Ibata}, {Ibata}, {Lewis}, {Martin}, {Conn},
  {Elahi}, {Arias}, \& {Fernando}}]{Ibata2014}
{Ibata}, R.~A., {Ibata}, N.~G., {Lewis}, G.~F., {et~al.} 2014, \apjl, 784, L6

\bibitem[{{Ibata} {et~al.}(2013){Ibata}, {Lewis}, {Conn}, {Irwin},
  {McConnachie}, {Chapman}, {Collins}, {Fardal}, {Ferguson}, {Ibata}, {Mackey},
  {Martin}, {Navarro}, {Rich}, {Valls-Gabaud}, \& {Widrow}}]{Ibata2013}
{Ibata}, R.~A., {Lewis}, G.~F., {Conn}, A.~R., {et~al.} 2013, \nat, 493, 62

\bibitem[Kang et 
al.(2005)]{j} Kang, X., Mao, S., Gao, L., \& Jing, Y.~P.\ 2005, \aap, 437, 383 

\bibitem[Khandai et al.(2014)]{Khandai2014} Khandai, N., Di Matteo, 
T., Croft, R., et al.\ 2014, arXiv:1402.0888 

\bibitem[{{Kroupa} {et~al.}(2010){Kroupa}, {Famaey}, {de Boer},
  {Dabringhausen}, {Pawlowski}, {Boily}, {Jerjen}, {Forbes}, {Hensler}, \&
  {Metz}}]{Kroupa2010}
{Kroupa}, P., {Famaey}, B., {de Boer}, K.~S., {et~al.} 2010, \aap, 523, A32

\bibitem[{{Kroupa} {et~al.}(2005){Kroupa}, {Theis}, \& {Boily}}]{Kroupa2005}
{Kroupa}, P., {Theis}, C., \& {Boily}, C.~M. 2005, \aap, 431, 517

\bibitem[{{Larson} {et~al.}(2011){Larson}, {Dunkley}, {Hinshaw}, {Komatsu},
  {Nolta}, {Bennett}, {Gold}, {Halpern}, {Hill}, {Jarosik}, {Kogut}, {Limon},
  {Meyer}, {Odegard}, {Page}, {Smith}, {Spergel}, {Tucker}, {Weiland},
  {Wollack}, \& {Wright}}]{Larson2011}
{Larson}, D., {Dunkley}, J., {Hinshaw}, G., {et~al.} 2011, \apjs, 192, 16

\bibitem[{{Li} \& {Helmi}(2008)}]{Li2008}
{Li}, Y.-S. \& {Helmi}, A. 2008, \mnras, 385, 1365

\bibitem[{{Libeskind} {et~al.}(2009){Libeskind}, {Frenk}, {Cole}, {Jenkins}, \&
  {Helly}}]{Libeskind2009}
{Libeskind}, N.~I., {Frenk}, C.~S., {Cole}, S., {Jenkins}, A., \& {Helly},
  J.~C. 2009, \mnras, 399, 550

\bibitem[{{Lovell} {et~al.}(2011){Lovell}, {Eke}, {Frenk}, \&
  {Jenkins}}]{Lovell2011}
{Lovell}, M.~R., {Eke}, V.~R., {Frenk}, C.~S., \& {Jenkins}, A. 2011, \mnras,
  413, 3013

\bibitem[{{Lynden-Bell}(1976)}]{LyndenBell1976}
{Lynden-Bell}, D. 1976, \mnras, 174, 695

\bibitem[{{McConnachie}(2012)}]{McConnachie2012}
{McConnachie}, A.~W. 2012, \aj, 144, 4

\bibitem[{{Metz} {et~al.}(2009){Metz}, {Kroupa}, {Theis}, {Hensler}, \&
  {Jerjen}}]{Metz2009}
{Metz}, M., {Kroupa}, P., {Theis}, C., {Hensler}, G., \& {Jerjen}, H. 2009,
  \apj, 697, 269

\bibitem[{{Pawlowski} \& {Kroupa}(2013)}]{Pawlowski2013b}
{Pawlowski}, M.~S. \& {Kroupa}, P. 2013, \mnras, 435, 2116

\bibitem[{{Pawlowski} {et~al.}(2012{\natexlab{a}}){Pawlowski}, {Kroupa},
  {Angus}, {de Boer}, {Famaey}, \& {Hensler}}]{Pawlowski2012b}
{Pawlowski}, M.~S., {Kroupa}, P., {Angus}, G., {et~al.} 2012{\natexlab{a}},
  \mnras, 424, 80

\bibitem[{{Pawlowski} {et~al.}(2011){Pawlowski}, {Kroupa}, \& {de
  Boer}}]{Pawlowski2011}
{Pawlowski}, M.~S., {Kroupa}, P., \& {de Boer}, K.~S. 2011, \aap, 532, A118

\bibitem[{{Pawlowski} {et~al.}(2013){Pawlowski}, {Kroupa}, \&
  {Jerjen}}]{Pawlowski2013a}
{Pawlowski}, M.~S., {Kroupa}, P., \& {Jerjen}, H. 2013, \mnras, 435, 1928

\bibitem[Pawlowski et al.(2014)]{Pawlowski2014b} Pawlowski, M.~S., 
Famaey, B., Jerjen, H., et al.\ 2014, arXiv:1406.1799 

\bibitem[Pawlowski 
\& McGaugh(2014)]{Pawlowski2014} Pawlowski, M.~S., \& McGaugh, S.~S.\ 2014, \mnras, 440, 908 

\bibitem[{{Pawlowski} {et~al.}(2012{\natexlab{b}}){Pawlowski},
  {Pflamm-Altenburg}, \& {Kroupa}}]{Pawlowski2012a}
{Pawlowski}, M.~S., {Pflamm-Altenburg}, J., \& {Kroupa}, P. 2012{\natexlab{b}},
  \mnras, 423, 1109

\bibitem[Vogelsberger et al.(2014)]{Vogelsberger2014} Vogelsberger, M., 
Genel, S., Springel, V., et al.\ 2014, \nat, 509, 177 

\bibitem[{{Wang} {et~al.}(2013){Wang}, {Frenk}, \& {Cooper}}]{Wang2013}
{Wang}, J., {Frenk}, C.~S., \& {Cooper}, A.~P. 2013, \mnras, 429, 1502

\bibitem[Yang et al.(2014)]{Yang2014} Yang, Y., Hammer, F., 
Fouquet, S., et al.\ 2014, arXiv:1405.2071 

\bibitem[Zentner et al.(2005)]{Zentner2005} Zentner, A.~R., 
Kravtsov, A.~V., Gnedin, O.~Y., \& Klypin, A.~A.\ 2005, \apj, 629, 219 

\end{thebibliography}
\end{document}